\documentclass[aps,prb,twocolumn,floatfix,showpacs,citeautoscript,longbibliography,superscriptaddress,hyperlinks]{revtex4-2}
\usepackage{amsmath}
\usepackage{graphicx,graphics}
\usepackage{epstopdf}
\usepackage{amsmath}
\usepackage{amssymb}
\usepackage{amsfonts}
\usepackage[colorlinks=true,citecolor=blue]{hyperref}
\usepackage{etoolbox}
\usepackage{xcolor}
\apptocmd{\sloppy}{\hbadness 10000\relax}{}{}

\newcommand{\revision}[1]{{{#1}}}

\begin{document}
\title{Lee-Yang theory of the superradiant phase transition in the open Dicke model}
\author{Fredrik Brange}
\affiliation{Department of Applied Physics, Aalto University, 00076 Aalto, Finland}
\author{Neill Lambert}
\affiliation{Theoretical Quantum Physics Laboratory, Cluster for Pioneering Research, RIKEN, Wakoshi, Saitama 351-0198, Japan}
\affiliation{RIKEN Center for Quantum Computing, Wakoshi, Saitama 351-0198, Japan}
\author{Franco Nori}
\affiliation{Theoretical Quantum Physics Laboratory, Cluster for Pioneering Research, RIKEN, Wakoshi, Saitama 351-0198, Japan}
\affiliation{RIKEN Center for Quantum Computing, Wakoshi, Saitama 351-0198, Japan}
\affiliation{Physics Department, University of Michigan, Ann Arbor, MI 48109-1040, USA}
\author{Christian Flindt}
\affiliation{Department of Applied Physics, Aalto University, 00076 Aalto, Finland}
\affiliation{RIKEN Center for Quantum Computing, Wakoshi, Saitama 351-0198, Japan}

\begin{abstract}
The Dicke model describes an ensemble of two-level atoms that are coupled to a confined light mode of an optical cavity. Above a critical coupling, the cavity becomes macroscopically occupied, and the system enters the superradiant phase. This phase transition can be observed by detecting the photons that are emitted from the cavity; however, it only becomes apparent in the limit of long observation times, while actual experiments are of a finite duration. To circumvent this problem, we here make use of recent advances in Lee-Yang theories of phase transitions to show that the superradiant phase transition can be inferred from the factorial cumulants of the photon emission statistics obtained during a finite measurement time. Specifically, from the factorial cumulants, we can determine the complex singularities of generating functions that describe the photon emission statistics, and by extrapolating their positions to the long-time limit, one can detect the superradiant phase transition. We also show that the convergence points determine the tails of the large-deviation statistics of the photon current. Our work demonstrates how phase transitions in the Dicke model and in other quantum many-body systems can be detected from measurements of a finite duration.
\end{abstract}

\maketitle

\section{Introduction}

\begin{figure}[b!]
    \centering
    \includegraphics[width=0.48\textwidth]{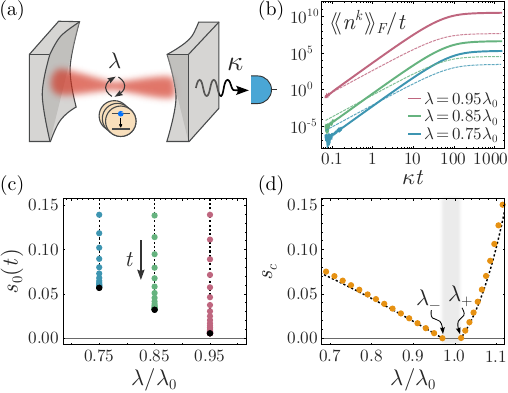}
    \caption{Lee-Yang theory of the open Dicke model. (a)~The system consists of an optical cavity with resonance frequency~$\omega_c$ that is  coupled to $N$ two-level atoms with energy splitting~$\hbar\omega_a$. The coupling amplitude is denoted by~$\lambda$, while~$\kappa$ is the rate at which photons are emitted from the cavity. (b) Fourth (dashed, $k=4$) and fifth (solid, $k=5$) factorial cumulant of the photon emission statistics in the normal phase \revision{defined in Eqs.~(\ref{eq:FMGF}-\ref{Cumulants})}. (c) Poles of the factorial cumulant generating function extracted from the results in panel~(b) \revision{using Eq.~(\ref{eq:pole_extract})}. The black dots mark the convergence points in the long-time limit\revision{, $s_c$}. (d) Extracted convergence points as a function of the coupling. Parameters are $\omega_c = 2 \kappa$ and $\omega_a = 0.5 \kappa$, and $\lambda_0$ denotes the critical point of the closed Dicke model, while $\lambda_-$ and $\lambda_+$ are the critical points of the open Dicke model according to Eqs.~(\ref{Lower limit lambda}, \ref{Upper limit lambda}).}
    \label{Fig1}
\end{figure}

The Dicke model, or the Dicke-Hepp-Lieb model, describes a quantum many-body system that exhibits a superradiant phase transition at a critical light-matter coupling~\cite{dicke, Hepp1973,Wang1973,Kirton2018}. It consists of a large ensemble of two-level atoms that interact with a single light-mode of an optical cavity, as illustrated in Fig.~\ref{Fig1}(a). It can also be realized using superconducting circuits~\cite{lambert09, Nataf2010,oliver11,Wang2014,lambert16, bamba16,Minganti:2021,rabl18,DiStefano2019,sahel}, trapped ions~\cite{genway14}, or collective electronic systems~\cite{simone}. Above the critical coupling, the system enters the superradiant phase, where the ground state acquires a macroscopic cavity occupation with the photon number being on the order of the number of atoms. Because of its simplicity, the model has been used to investigate a range of critical phenomena, including  chaos~\cite{Emary2003,PhysRevE.67.066203} and entanglement~\cite{Lambert2004} at criticality. Still, a direct experimental observation of the superradiant phase transition has proven challenging~\cite{Kirton2018}.

To observe the superradiant phase transition, effective physical realizations involving driven-dissipative systems have been explored. In these setups, internal \cite{carmichaelPRA, Zhiqiang:17} or motional \cite{Baumann2010,nagyPRL} degrees of freedom are pumped to realize effective strong couplings, but 
with finite dissipation rates. The combination of the external driving and dissipation results in a non-equilibrium phase transition, which is of a different universality class than for the closed system~\cite{carmichaelPRA, dallaPRA, Kirton2018,Carollo:2021,Boneberg:2022}. However, this phase transition is easier to access experimentally, and it will be the focus of our work. The phase transition occurs in the thermodynamic limit of many atoms and is visible in the photon emission statistics collected over a long time. For example, the cavity occupation changes abruptly at the critical coupling~\cite{Chen2024}. At the same \revision{time}, however, measurements of the photon counting statistics are of a finite duration, so practical approaches to predict the long-time behavior from finite-time statistics are needed.

In this work, we make use of recent advances in Lee-Yang theories of phase transitions to observe the superradiant phase transition in the open Dicke model from the photon counting statistics measured during a finite observation time. The theory of phase transitions by Lee and Yang was originally formulated for equilibrium phase transitions~\cite{Yang:1952,Lee:1952,Blythe2003,Bena2005}. It concerns the zeros of the partition function in the complex plane of a control parameter, for example, the inverse temperature or an external magnetic field. For systems of finite size, the zeros are complex. However, if the system exhibits a phase transition, the zeros will move towards the critical point on the real axis as the thermodynamic limit is approached. The theory of Lee and Yang has been extended to other types of phase transitions, including non-equilibrium phase transitions~\cite{Blythe2002,Blythe2003,Bena2005}, trajectory phase transitions~\cite{Flindt2013}, dynamical quantum phase transitions~\cite{Heyl:2013,Xu2020,Peotta2021,Brange:2022}, and quantum phase transitions at zero temperature~\cite{Kist2021,Vecsei2022,Vecsei:2023}. Lee-Yang zeros have also been determined in several experiments~\cite{Binek1998,Peng2015,Brandner2017,gao:2024}.

In the approach we follow here, the phase transition is inferred from the photon counting statistics collected during a finite observation time~\cite{Flindt2013,Brandner2017}. As illustrated in Fig.~\ref{Fig1}(b), we use the high factorial cumulants of the photon counting statistics to extract the dominant pole of the corresponding generating function, which plays the role of the partition function for equilibrium systems.  As shown in Fig.~\ref{Fig1}(c), we can then determine the convergence point in the limit of long observation times. Away from the phase transition, the convergence point is non-zero. However, as we tune the coupling to its critical value, the dominant pole converges to zero, and we can detect the phase transition, as illustrated in Fig.~\ref{Fig1}(d).

The rest of our article is organized as follows. In Sec.~\ref{sec:dicke_model}, we derive a master equation for the open Dicke model in the limit of many atoms. Here, we follow  earlier works on the Dicke model, and we include this section for the sake of completeness. In Sec.~\ref{sec:FCS}, we consider the photon emission statistics from the cavity and derive an analytic expression for the generating function, which is valid at all times. In Sec.~\ref{sec:LY}, we describe the Lee-Yang theory to detect the superradiant phase transition from the photon counting statistics measured during a finite observation time. In Sec.~\ref{sec:PT}, we use this approach to determine the critical behavior of the open Dicke model from the high factorial cumulants of the photon counting statistics. We also show how the extracted convergence points control the tails of the large-deviation statistics of the photon current. Finally, in Sec.~\ref{sec:conclusion}, we conclude on our work and provide an outlook on possible developments for the future. A few technical details are deferred to App.~\ref{AppA}.

\section{Dicke model}
\label{sec:dicke_model}

Figure~\ref{Fig1}(a) shows a physical implementation of the Dicke model consisting of a single-mode optical cavity coupled to an ensemble of two-level atoms, \revision{noting that other physical realizations are also possible}~\cite{lambert09,Nataf2010,oliver11,Wang2014,lambert16,bamba16, Minganti:2021,rabl18,DiStefano2019,sahel,genway14,simone}. Some implementations are based on driving internal transitions of multi-level atoms \cite{carmichaelPRA, Zhiqiang:17}, or on the motional states of atoms in a Bose-Einstein condensate \cite{Baumann2010,nagyPRL}. The latter implementation involves dissipative effects, which we will also include in our treatment.  

We denote the frequency of the cavity by $\omega_c$, while $\omega_a$ determines the energy splitting of the  two-level atoms. The Hamiltonian of the cavity-atom system reads
\begin{equation}
\hat H = \hbar \omega_c \hat c^\dagger \hat c + \hbar \omega_a\hat J_z + \frac{\hbar \lambda}{\sqrt{N}}\left( \hat c^\dagger +\hat c\right)\left(\hat J_+ +\hat J_-\right),
\label{Hamiltonian}
\end{equation}
where $N$ is the number of atoms, and $\lambda$ denotes the coupling between the cavity and the atoms. We have also introduced the creation and annihilation operators of the cavity,  $\hat c^\dagger$ and $\hat c$. We treat the ensemble of two-level atoms as a pseudo-spin of length $N/2$, and  $\hat J_z$ and $\hat J_\pm$ are then collective angular momentum operators with $\hat J_z$ measuring the angular momentum in the $z$-direction, while $\hat J_\pm$ are the usual raising and lowering operators.

As illustrated in Fig.~\ref{Fig1}(a), photons are emitted from the cavity at the rate $\kappa$. Correspondingly, the density matrix of the open cavity-atom system evolves according to the the Lindblad equation
\begin{equation}
\frac{d}{dt} \hat \rho(t) = -\frac{i}{\hbar}\left[ \hat H,\hat \rho(t)\right] +\kappa \mathcal{D}[\hat c]\hat \rho(t),
\label{Lindblad equation}
\end{equation}
where the commutator accounts for the unitary dynamics of the system if isolated.   The dissipator
\begin{equation}
\mathcal{D}[\hat c]\hat \rho(t) = \hat c\hat \rho(t) \hat c^\dagger- \frac{1}{2} \left\{ \hat c^\dagger \hat c, \hat \rho(t)\right\}
\end{equation}
describes the emission of photons from the cavity. Related to the realizations discussed in Refs.~\cite{carmichaelPRA,Baumann2010}, the cavity and atomic frequencies in Eq.~(\ref{Hamiltonian}) are effective frequencies in a rotating frame of a drive, and  the dissipation in Eq.~(\ref{Lindblad equation}) acts locally on the cavity mode only. In that respect, it is a non-equilibrium problem as the dissipation will not cool the system to its ground state, but instead it will lead to a non-equilibrium steady-state. 

We now derive an effective model in the thermodynamic limit of many atoms. To this end, we follow Refs.~\cite{PhysRevE.67.066203,PhysRevA.87.043840} and repeat the main steps for the sake of completeness. First, we employ a Holstein-Primakoff transformation \cite{PhysRev.58.1098,PhysRevB.44.2227} to represent the many atoms by a single bosonic mode. Specifically, we express the angular momentum operators in terms of bosonic operators as
\begin{equation}
\begin{split}
\hat J_+ =& \hat a^\dagger \sqrt{N-\hat a^\dagger \hat a},\\
\hat J_- =& \sqrt{N-\hat a^\dagger \hat a}\hat a, 
\end{split}
\end{equation}
and
\begin{equation}
\hat J_z = \hat a^\dagger\hat a -N/2
\end{equation}
where the creation and annihilation operators fulfill the usual bosonic commutation relation $[\hat a,\hat a^\dagger]=1$.
Under this transformation, the Hamiltonian becomes
\begin{equation}
\begin{split}
\hat H =&\hbar \omega_c \hat c^\dagger \hat c+ \hbar \omega_a\hat a^\dagger \hat a\\
+&\hbar \lambda \left(\hat c^\dagger +\hat c\right)\left(\hat a^\dagger \sqrt{1-\frac{\hat a^\dagger \hat a}{N}} + \sqrt{1-\frac{\hat a^\dagger \hat a}{N}} \hat a \right),
\label{Holstein-Primakoff transformed Hamiltonian}
\end{split}
\end{equation}
having omitted an offset that is proportional to $N$.

We now consider the thermodynamic limit. For small couplings, the system is in the normal phase, where averages of $\hat a^\dagger\hat a$ do not grow with $N$, and we obtain
\begin{equation}
\hat H_{\mathrm{NP}} =\hbar \omega_c \hat c^\dagger \hat c+ \hbar \omega_a \hat a^\dagger \hat a +\hbar \lambda \left( \hat c^\dagger + \hat c\right) \left(\hat a^\dagger +\hat a \right).
\label{Effective Hamiltonian N}
\end{equation}
The dissipator in Eq.~\eqref{Lindblad equation} is independent of the number of atoms. 
Therefore, the Lindblad equation becomes
\begin{equation}
\frac{d}{dt} \hat \rho(t) = -\frac{i}{\hbar}\left[ \hat H_\mathrm{NP},\hat \rho(t)\right] +\kappa \mathcal{D}[\hat c]\hat \rho(t).
\end{equation}
Next, we diagonalize the Hamiltonian using eigenvalue decompositions such that
\begin{equation}
\hat c = \varepsilon_{1_-} \hat d_1 +\varepsilon_{1_+} \hat d_1^\dagger + \varepsilon_{2_-}\hat d_2+\varepsilon_{2_+}\hat d_2^\dagger
    \label{Diagonalization}
\end{equation}
with a similar expression for the operator $\hat a$, which eventually drops out. The Hamiltonian then becomes
\begin{equation}
    \hat H_{D} = \hbar\omega_1 \hat d_1^\dagger \hat d_1 +\hbar\omega_2 \hat d_2^\dagger \hat d_2
    \label{Eigenvalue decomposed Hamiltonian}
\end{equation}
with the corresponding master equation reading
\begin{equation}
\begin{split}
    &\frac{d}{dt}\hat \rho(t) = -\frac{i}{\hbar} \left[\hat H_D,\hat \rho(t) \right]  \\
    &+\kappa\left(\varepsilon_{1_-}^2 \mathcal{D}[\hat d_1]+\varepsilon_{1_+}^2 \mathcal{D}[\hat d_1^\dagger]+\varepsilon_{2_-}^2 \mathcal{D}[\hat d_2]+\varepsilon_{2_+}^2 \mathcal{D}[\hat d_2^\dagger]\right)\hat \rho(t),
\end{split}
\label{eq:finalGME}
\end{equation}
having omitted fast-rotating terms of the form $\hat d_1^\dagger \hat d_2^\dagger$. The coefficients $\varepsilon_{1_\pm,2_\pm}$ and the frequencies $\omega_{1,2}$ are defined in App.~\ref{AppA}. The smallest frequency vanishes if
\begin{equation}
    \lambda_- = \frac{\sqrt{\omega_c\omega_a}}{2},
    \label{Lower limit lambda}
\end{equation}
which sets an upper limit on $\lambda$ for the normal phase.  

In the superradiant phase, the average occupation of the cavity and the number of excited atoms increase with~$N$. We thus employ a mean-field ansatz to find an effective Hamiltonian in the thermodynamic limit. To this end, we define the operators
\begin{equation}
\check c = \hat c -\sqrt{r_cN}, \qquad \check a = \hat a +\sqrt{r_aN},
\label{Mean-field operators}
\end{equation}
which are independent of $N$, and $r_a$ is a real constant, while~$r_c$ can be complex. The macroscopic occupation is determined such that all terms proportional to $\sqrt{N}$ vanish in the master equation, yielding the condition for a stable stationary state that~\cite{PhysRevE.67.066203,PhysRevA.87.043840}
\begin{equation}
r_c =  \frac{4\lambda^2 \left[1-\left(\lambda_0/\lambda \right)^4 \right]}{(2\omega_c-i\kappa)^2},
\end{equation}
and
\begin{equation}
r_a =[1-\left( \lambda_0/\lambda\right)^2]/2,.
\label{alpha and beta}
\end{equation}
Here, we have defined the coupling
\begin{equation}
 \lambda_0 = \sqrt{(\kappa^2+4\omega_c^2)\omega_a/16\omega_c},  
\end{equation}
above which occupations become macroscopic in the limiting case of vanishing emission rate, $\kappa=0$.

The Hamiltonian in the superradiant phase now reads
\begin{equation}
\begin{split}
    \hat H_\mathrm{SP} = \hbar \omega_c \check c^\dagger \check c&+\hbar\Omega_a \check a^\dagger \check a +\hbar \lambda_{\mathrm{aa}}\left(\check a^\dagger + \check a \right)^2 \\
+&  \hbar \lambda_{\mathrm{ca}} \left(\check c^\dagger + \check c \right) \left( \check a^\dagger + \check a\right),
\end{split}
\end{equation}
where we have introduced the frequency
\begin{equation}
\Omega_a = \omega_a+\lambda\sqrt{N}\frac{2r_a}{\sqrt{1-r_a}}\mathrm{Re}\{\sqrt{r_c}\},
\end{equation}
together with the couplings
\begin{equation}
\lambda_{\mathrm{ca}} = \lambda \frac{1-2r_a}{\sqrt{1-r_a}},
\end{equation}
and
\begin{equation}
\lambda_{\mathrm{aa}} = \lambda\frac{(2-r_a)\sqrt{r_a}}{2(1-r_a )^{3/2}}\mathrm{Re}\{\sqrt{r_c}\}.
\end{equation}
Moreover, the master equation becomes
\begin{equation}
\frac{d}{dt} \hat \rho(t) = -\frac{i}{\hbar}\left[ \hat H_\mathrm{SP},\hat \rho(t)\right] +\kappa \mathcal{D}[\check c]\hat \rho(t).
\end{equation}
Just as in the normal phase, we can diagonalize the Hamiltonian using eigenvalue decompositions. We then arrive at the same Hamiltonian and master equation as in Eqs.~(\ref{Eigenvalue decomposed Hamiltonian}) and (\ref{eq:finalGME}), however, with different coefficients~$\varepsilon_{1_\pm,2_\pm}$ and  frequencies~$\omega_{1,2}$ as detailed in App.~\ref{AppA}. In this case, the smallest frequency vanishes for
\begin{equation}
    \lambda_+ = \frac{\left(\kappa^2+4\omega_c^2 \right)^{3/4}\sqrt{\omega_a}}{4\sqrt{2}\omega_c},
    \label{Upper limit lambda}
\end{equation}
which sets a lower limit on $\lambda$ for the superradiant phase. We note that $\lambda_+$ depends on $\kappa$, and only when $\kappa \ll \omega_c$, we have $\lambda_+ = \lambda_-$. It will also be important that the master equation in Eq.~(\ref{eq:finalGME}) describes two uncoupled quantum harmonic oscillators.
 
\section{Photon emission statistics}
\label{sec:FCS}
To observe the  phase transitions using our Lee-Yang theory, we consider the probability $P(n,t)$ that $n$ photons are emitted during the time span $[0,t]$. It is useful to introduce the factorial moment generating function
\begin{equation}
\mathcal{M}_F(s,t) = \sum_{n=0}^\infty P(n,t)(1+s)^n,
\label{eq:FMGF}
\end{equation}
which yields the factorial moments upon differentiation with respect to the counting variable~$s$ as
\begin{equation}
    \langle n^k\rangle_F(t) = \partial^k_s \mathcal{M}_F(s,t)\big|_{s=0}.
\end{equation}
The factorial moments are given by the ordinary ones as
\begin{equation}
    \langle n^k\rangle_F = \langle n(n-1)...(n-k+1)\rangle. 
\end{equation}
For instance, the first two factorial moments read $\langle n\rangle_F = \langle n\rangle$ and $\langle n^2\rangle_F = \langle n^2\rangle - \langle n \rangle$ in terms of the ordinary ones. We also define the factorial cumulant generating function
\begin{equation}
\mathcal{F}_F(s,t)= \ln \mathcal{M}_F(s,t),
\end{equation}
which yields the factorial cumulants as
\begin{equation}
\langle\!\langle n^k\rangle\!\rangle_F(t)= \partial^k_s  \mathcal{F}_F(s,t)\big |_{s=0}.
\label{Cumulants}
\end{equation}
The factorial cumulants can also be expressed in terms of the ordinary ones as
\begin{equation}
    \langle\!\langle n^k\rangle\!\rangle_F = \langle\!\langle n(n-1)...(n-k+1)\rangle\!\rangle. 
\end{equation}
Factorial cumulants are useful to characterize discrete quantities, such as the number of photons~\cite{Kambly:2011,Kambly:2013,Stegmann:2015,Konig:2021,Kleinherbers:2022}. Indeed, for a Poisson distribution, only the first factorial cumulant is non-zero. By contrast, ordinary cumulants are useful to characterize continuous variables, and they are defined so that only the first two cumulants are non-zero for a Gaussian distribution. 

To find the generating functions, we resolve the density matrix with respect to the number of emitted photons during the time span $[0,t]$, which defines $\hat \rho(n,t)$~\cite{Plenio:1998}. We can then express the photon counting statistics as
\begin{equation}
P(n,t) = \text{tr}\left\{ \hat \rho(n,t) \right\}.
\end{equation}
These density matrices obey the system of equations
\begin{equation}
\begin{split}
    \frac{d}{dt}\hat \rho(n,t) &=  -\frac{i}{\hbar}\left[\hat H,\hat \rho(n,t) \right]\\
    &+ \kappa \left(\hat c \hat \rho(n-1,t) \hat c^\dagger -\frac{1}{2} \left\{\hat c^\dagger \hat c,\hat \rho(n,t) \right\} \right).
\end{split}    
\end{equation}
To solve these coupled equations, we define
\begin{equation}
    \hat\rho(s,t) = \sum_{n = 0}^\infty \hat \rho(n,t) (1+s)^n,
\end{equation}
whose equation of motions reads
\begin{equation}
    \frac{d}{dt}\hat \rho(s,t) = -\frac{i}{\hbar}\left[\hat H,\hat \rho(s,t) \right]+  \kappa \mathcal{D}_s[\hat c] \hat \rho(s,t),
\end{equation}
where the counting variable now enters the dissipator as
\begin{equation}
\mathcal{D}_s[\hat c]\hat \rho = (1+s)\hat c \hat \rho \hat c^\dagger -\frac{1}{2} \left\{\hat c^\dagger \hat c,\hat \rho \right\}.
\end{equation}
Solving for $\rho(s,t)$, we obtain the generating function as
\begin{equation}
\mathcal{M}_F(s,t) = \text{tr}\left\{\hat \rho(s,t) \right\}. 
\end{equation}

Next, we employ the diagonalization in Eq.~\eqref{Diagonalization} and switch to the interaction picture with respect to the Hamiltonian that governs the unitary dynamics. In the rotating-wave approximation, we neglect cross-terms such as $\hat d_1^\dagger\hat d_2^\dagger$, and we thus arrive at the master equation 
\begin{equation}
\begin{split}
\frac{d}{dt}&\tilde{\hat \rho}(s,t) = \kappa\bigg(\varepsilon_{1_-}^2 \mathcal{D}_s[\hat d_1]+\varepsilon_{1_+}^2  \mathcal{D}_s[\hat d_1^\dagger]\\
    &+\varepsilon_{2_-}^2  \mathcal{D}_s[\hat d_2]+\varepsilon_{2_+}^2  \mathcal{D}_s[\hat d_2^\dagger] + N |r_c| s\bigg)\tilde{\hat \rho}(s,t),
\end{split}
\end{equation}
where $r_c$ in the superradiant phase is given by Eq.~\eqref{alpha and beta}, while it vanishes in the normal phase. 

The master equation above describes three independent processes. There are the photon emissions from two independent quantum harmonic oscillators as well as the photon emission from a Poisson process with rate $\kappa N |r_c| $. As a result, the generating function factorizes as  
\begin{equation}
\mathcal{M}_F(s,t) = \mathcal{M}^{(1)}_F(s,t)\mathcal{M}^{(2)}_F(s,t)\mathcal{M}^{(p)}_F(s,t), 
\label{MGF}
\end{equation}
where $\mathcal{M}^{(1,2)}_F(s,t)$ are the generating functions corresponding to each of the two harmonic oscillators, while
\begin{equation}
\mathcal{M}^{(p)}_F(s,t)=\exp{(\kappa N |r_c| t s)}
\end{equation}
describes the Poissonian emission of photons. In a recent work, some of us determined the generating function of a  quantum harmonic oscillator, which reads~\cite{PhysRevB.99.085418} 
\begin{equation}
\mathcal{M}^{(j)}_F(s,t)=\frac{\xi^{(2)}_j e^{\kappa_j t}}{\xi^{(2)}_j\cosh (\xi^{(2)}_j\kappa_j t)+[\xi^{(1)}_j]^2 \sinh(\xi_j^{(2)}\kappa_j t)},
\end{equation}
where we have defined the constants
\begin{equation}
\begin{split}
\kappa_1 &= \kappa \left(\varepsilon_{1_-}^2 - \varepsilon_{1_+}^2 \right)/2,\\  \kappa_2 &= \kappa \left(\varepsilon_{2_-}^2 - \varepsilon_{2_+}^2 \right)/2,
\end{split}
\end{equation}
together with the functions
\begin{equation}
\xi_{j}^{(k)} = \sqrt{1-4 \bar n_j(1+\bar n_j)[(1+s)^k-1]}
\end{equation}
with
\begin{equation}
\begin{split}
\bar n_1 &= \varepsilon_{1_+}^2/(\varepsilon_{1_-}^2 -\varepsilon_{1_+}^2),\\  \bar n_2 &= \varepsilon_{2_+}^2/(\varepsilon_{2_-}^2 -\varepsilon_{2_+}^2).
\end{split}
\end{equation}
These expressions fully characterize the  emission statistics of the  Dicke model for all possible observation times.

\section{Lee-Yang theory}
\label{sec:LY}

\begin{figure*}
    \centering
    \includegraphics[width=0.98\textwidth]{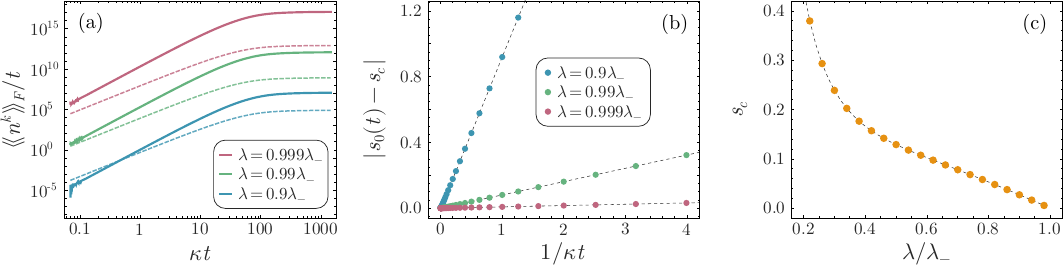}
      \caption{Lee-Yang theory of the normal phase. (a) Fourth (dashed) and fifth (solid) factorial as a function of the observation time in the normal phase. (b) The dominant pole extracted  from the factorial cumulants in panel (a). (c) Convergence point of the dominant pole in the long-time limit as  a function of the coupling. Parameters are $\omega_c=2 \kappa$ and $\omega_a = 0.5 \kappa$.}
      \label{Fig2}
\end{figure*}

We are now ready to use our Lee-Yang theory to detect the superradiant phase transition. In the original theory of equilibrium phase transitions by Lee and Yang, they considered the zeros of the partition function in the complex plane of the control parameters; for instance, an external magnetic field or the inverse temperature~\cite{Yang:1952,Lee:1952,Blythe2003,Bena2005}. For systems of finite size, the zeros are complex and the imaginary part is finite. However, if the system exhibits a phase transition, the zeros will approach the point on the real-axis, where the phase transition occurs, as the system size is increased. The Lee-Yang theory of equilibrium phase transitions has found use across a wide range of fields in physics and related areas~\cite{Blythe2003,Bena2005}. It has also been shown that Lee-Yang zeros can be experimentally determined from the fluctuations of the observable that couples to the control parameter~\cite{Deger2018}, for instance, the inverse temperature couples to the  energy~\cite{Brange:2023}, and a magnetic field may couple to the magnetization of a spin system~\cite{Wei:2012,Wei:2014,Deger2020}. In addition, Lee-Yang zeros have been determined in several experiments~\cite{Binek1998,Peng2015,Brandner2017,gao:2024}.

\revision{While the theory of Lee and Yang  was developed for equilibrium phase transitions~\cite{Yang:1952,Lee:1952}, it has subsequently been extended to several types of non-equilibrium situations}~\cite{Blythe2002,Blythe2003,Bena2005,Flindt2013,Heyl:2013,Xu2020,Peotta2021,Brange:2022,Kist2021,Vecsei2022,Vecsei:2023}. In such nonequilibrium situations, the phase transitions are not manifested in the partition function or the associated free-energy density. Instead, other quantities, such as the moment generating function or the Loschmidt amplitude, play the role of the partition function, whose complex zeros determine the phase behavior of the system. In the approach that we follow here, the generating function in Eq.~(\ref{eq:FMGF}) plays the role of the partition function. The partition function is given by a sum over all possible microconfigurations of an equilibrium system weighted by Boltzmann factors. Similarly, the factorial moment generating function is given by a sum over all possible dynamical trajectories, characterized by the number $n=0,1,2\ldots$ of photons that have been emitted during the time span $[0,t]$, weighted by the counting variable $s$. For an equilibrium system, the thermodynamic limit is approached as the system size, for example, the number of spins in a spin lattice,  is increased. By contrast, in our case, the thermodynamic limit is approached as the observation time is increased, and the quantum jump trajectories become long. Thus, as our dynamical free-energy density, we take the scaled factorial cumulant generating function, defined in the long-time limit as
\begin{equation}
    \Theta_F(s)=\lim_{t\rightarrow \infty} \mathcal{F}_F(s,t)/t= \lim_{t\rightarrow \infty} \ln\{\mathcal{M}_F(s,t)\}/t.
\end{equation}
In this case, the system exhibits a phase transition, if the dynamical free-energy density becomes nonanalytic at $s=0$. However, the nonanalytic behavior only emerges in the long-time limit, and one may wonder, if it can be observed in an experiment, where the observation time remains finite. To this end, we note that the nonanalytic behavior of the dynamical free-energy at $s=0$ are due to zeros or poles of the factorial moment generating function that approach the origin as the observation time increases. As we will see, these zeros and poles can be determined from measurements of the factorial moments at finite times, and  one can determine their converge points in the limit of long observation times by extrapolation.

\begin{figure*}[t]
    \centering
    \includegraphics[width=0.98\textwidth]{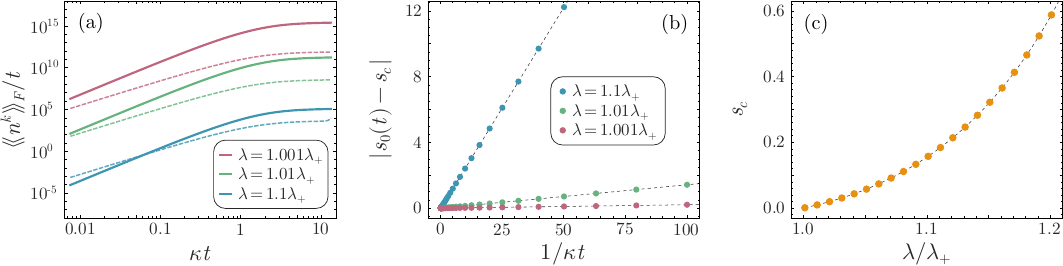}
      \caption{Lee-Yang theory of the superradiant phase. (a) Fourth (dashed) and fifth (solid) factorial cumulant as a function of the observation time in the superradiant phase. (b) The dominant pole extracted from the results in panel (a). (c) Convergence point in the long-time limit as  a function of the coupling. Parameters are $N=100$, $\omega_c=2 \kappa$, and $\omega_a = 0.5 \kappa$.}
      \label{Fig3}
\end{figure*}

In general, the factorial moment generating function can have both zeros and poles in the complex plane of the counting variable. For systems of finite size, the partition function only has zeros. However, when dealing with bosons for example, infinitely many states are involved, and the partition function may also have poles~\cite{Brange:2023}. In earlier works, we have developed methods that allow us to extract both the zeros and the poles from the high cumulants~\cite{Peotta2021,Brange:2023}. However, for the factorial moment generating function in Eq.~(\ref{MGF}), the situation simplifies since it only has poles. (At the points, where the numerator vanishes, $\xi_i^{(2)} = 0$, the denominator also vanishes, and the function is non-zero.) Thus, in the following, we formally expand the factorial moment generating function in terms of its unknown poles, $s_q(t)$, as
\begin{equation}
\mathcal{M}_F(s,t) = e^{s c} \prod_{q=0}^\infty \left[ 1-s/s_q(t)\right]^{-1}
\end{equation}
where $c$ is independent of $s$. Using the expression for the factorial cumulants in Eq.~\eqref{Cumulants}, we then have
\begin{equation}
\langle\!\langle n^k\rangle\!\rangle_F(t)= \sum_{q=0}^\infty \frac{(k-1)!}{s^k_q(t)},\quad  k> 1.
\end{equation}
From this expression, we see that the high factorial cumulants are dominated by the pole that is closest to the origin, $s=0$, which allows us to express them as
\begin{equation}
\langle\!\langle n^k\rangle\!\rangle_F(t)  \simeq \frac{(k-1)!}{s^k_0(t)}, \quad k\gg 1
\label{eq:pole_approx}
\end{equation}
for sufficiently high orders. Inverting this expression for the closest pole, we find that it can be expressed as 
\begin{equation}
s_0(t)  \simeq  (k-1) \frac{\langle\!\langle n^{k-1}\rangle\!\rangle_F(t)}{\langle\!\langle n^k\rangle\!\rangle_F(t)}, \quad k\gg 1
\label{eq:pole_extract}
\end{equation}
in terms of the high factorial cumulants. Thus, from two high factorial cumulants, which are measurable quantities, we can determine the pole that is closest to the origin, and we can follow its motion as a function of the observation time to determine its convergence point in the thermodynamic limit of long observation times.

\section{Phase transitions}
\label{sec:PT}

We now extract the pole that is closest to the origin of the complex plane of the counting variable from the factorial cumulants. This procedure is illustrated in Fig.~\ref{Fig2}(a), where we show the fourth and fifth factorial cumulants as a function of the observation time for three different couplings in the normal phase. \revision{We have checked that the results do not significantly change, if the cumulant order is increased, which shows that the approximations in Eqs.~(\ref{eq:pole_approx},\ref{eq:pole_extract}) are valid.} For each coupling, we extract the position of the dominant pole, which we show in Fig.~\ref{Fig2}(b) as a function of the inverse observation time. The position of the pole is well-approximated by the expression~\cite{Brandner2017,Deger2018,Deger2020,Brange:2023}
\revision{
\begin{equation}
    |s_0(t)-s_c|\propto 1/\kappa t,
\end{equation}
}where $s_c$ is the convergence point in the limit of long times. Thus, for each value of the coupling, we can extract the convergence point in the long-time limit, and in Fig.~\ref{Fig2}(c), we show the extracted convergence point as a function of the coupling. For small couplings, the convergence point does not reach the origin. However, as we approach the critical coupling, $\lambda\simeq \lambda_{-}$, the convergence point becomes smaller, and it eventually vanishes at $\lambda=\lambda_{-}$, signaling a phase transition. Thus, from measurements of the forth and fifth factorial cumulants at finite times, one can detect the phase transition at $\lambda=\lambda_{-}$. In particular, from Figs.~\ref{Fig2}(a,b), we see that one would be able to determine the convergence point in the long-time limit from measurements of the factorial cumulants for times that are shorter than $\kappa t=1$. \revision{Here, it should be noted that the extraction of the dominant pole in Eq.~(\ref{eq:pole_extract}) works best for short times, where the poles are well-separated. Also, high factorial cumulants can be more accurately measured at short times~\cite{Brandner2017,Kleinherbers:2022}.}

In Fig.~\ref{Fig3}, we take the same approach for large couplings, where the system is in the superradiant phase. In Fig.~\ref{Fig3}(a), we show the fourth and fifth factorial cumulants as a function of the observation time for different couplings. We then show the dominant pole as a function of the inverse observation time, which allows us to extract the convergence points in the long-time limit as shown in Fig.~\ref{Fig3}(b). In Fig.~\ref{Fig3}(c), we show the convergence point as a function of the coupling in the superradiant phase. For large couplings,  $\lambda> \lambda_{+}$, the convergence point does not reach zero. However, it becomes smaller as the coupling is reduced, and it finally reaches the origin of the complex plane for $\lambda=\lambda_{+}$. Thus, we again see that it is possible to detect the phase transition at $\lambda=\lambda_{+}$, which occurs in the long-time limit, from observations of the high factorial cumulants at finite times.

\begin{figure*}
    \centering
    \includegraphics[width=0.98\textwidth]{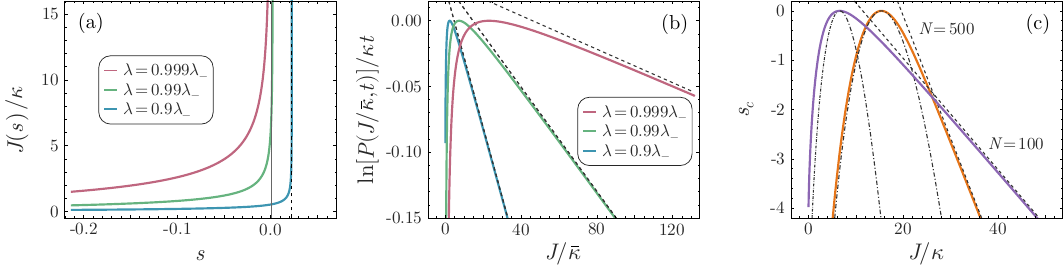}
    \caption{Large-deviation statistics of the photon emission current. (a) The $s$-biased current, $J(s) =\Theta'(s)$, in the normal phase given by  Eq.~\eqref{Analytic expression s-dependent current}. (b) Large-deviation  statistics in the normal phase given by Eq.~\eqref{Analytic expression large-deviation statistics}. The dashed lines are the approximation of the tails based on Eq.~\eqref{eq:LDFapprox}. (c) Large-deviation statistics in the superradiant phase with $\lambda = 1.1\lambda_+$ and $N=100$ and $N=500$. The other parameters are $\omega_c=2 \kappa$ and $\omega_a = 0.5 \kappa$. The dash-dotted lines represent Poisson distributions.}
    \label{Large deviation plots}
\end{figure*}

The results above show how we can detect phase transitions that occur in the limit of long times from measurements of the factorial cumulants during a finite duration. As we now go on to show, the extracted convergence points also influence the large-deviation statistics of the photon emission current. To this end, we invert the expression for the factorial moment generating function in Eq.~(\ref{eq:FMGF}) and write the probability distribution as
\begin{equation}
    P(n,t) = \frac{1}{2\pi i}\int_{-i\pi}^{i\pi} ds \mathcal{M}_F(e^s-1,t) e^{ -ns}.
\end{equation}
Now, defining the photon emission current as $J=n/t$ and considering the limit of long times, we can write
\begin{equation}
    P(J
    ,t) = \frac{1}{2\pi i}\int_{-i\pi}^{i\pi} ds \exp\{[\Theta(s) -s J]t\},
    \label{eq:P(n)}
\end{equation}
where we have used that $\Theta(s)=\Theta_F(e^s-1)$ is the scaled cumulant generating function. From Eq.~(\ref{MGF}), we find
\begin{equation}
\begin{split}
\Theta(s) =\sum_{j=1}^2 \kappa_j &\left( 1- \sqrt{1-4\bar n_j (1+\bar n_j) (e^{2s}-1)} \right)\\
&+\kappa N |r_c|(e^s-1)
\label{eq:scaledCGF}
\end{split}
\end{equation}
in agreement with Ref.~\cite{PhysRevA.87.043840}. At long times, the integral in Eq.~(\ref{eq:P(n)}) is amenable to a saddle-point approximation, such that the large-deviation statistics becomes~\cite{Touchette2009}
\begin{equation}
    \frac{\ln[P(J,t)]}{t} \simeq \Theta(s_{\mathrm{sp}}) -s_\mathrm{sp} J,
    \label{Large deviation statistics}
\end{equation}
where $s_\mathrm{sp} =s_\mathrm{sp}(J)$ solves the saddle-point equation
\begin{equation}
    \Theta'(s_\mathrm{sp}) = J.
\end{equation}

With these definitions at hand, we can calculate the large-deviation statistics of the photon emission current. In the normal phase, we note  that $\bar n_1 \gg \bar n_2$, if $\omega_c \gg \omega_a$ or $\omega_c \ll \omega_a$, and the term with $j=2$ in Eq.~\eqref{eq:scaledCGF} is then negligible. In that case, we find
\begin{equation}
\Theta'(s) = \frac{2 e^{2s} \varepsilon_{1_+}^2 \varepsilon_{1_-}^2 \kappa/(\varepsilon_{1_-}^2-\varepsilon_{1_+}^2 )}{\sqrt{1-(e^{2s}-1)[2\varepsilon_{1_+}\varepsilon_{1_-}/(\varepsilon_{1_-}^2-\varepsilon_{1_+}^2)]^2}},
\label{Analytic expression s-dependent current}
\end{equation}
which we will also refer to as the $s$-biased current and denote by $J(s) =\Theta'(s)$. In Fig.~\ref{Large deviation plots}(a), we show it for three different couplings in the normal phase. 

The solution to the saddle-point equation now becomes 
\begin{equation}
    s_\mathrm{sp} =\frac{1}{2}\ln\left[J^2\frac{\sqrt{1+\left(\frac{\bar \kappa}{J} \right)^2}-1}{2\varepsilon_{1_+}^2\varepsilon_{1_-}^2 \kappa^2} \right],
\end{equation}
where we have defined
\begin{equation}
\bar \kappa =(\varepsilon_{1_+}^2+\varepsilon_{1_-}^2)\kappa.
\end{equation}
In the normal phase, we then find the expression
\begin{widetext}
\begin{equation}
    \frac{\ln[P(J,t)]}{t} \simeq \frac{\varepsilon_{1_-}^2-\varepsilon_{1_+}^2}{2}\kappa -\frac{J}{2}\sqrt{\left(\frac{\bar \kappa}{J} \right)^2-2\left(\sqrt{1+\left(\frac{\bar \kappa}{J} \right)^2}-1 \right)}-\frac{J}{2} \ln\left[ J^2\frac{\sqrt{1+\left(\frac{\bar \kappa}{J} \right)^2}-1}{2\varepsilon_{1_+}^2\varepsilon_{1_-}^2 \kappa^2} \right],
    \label{Analytic expression large-deviation statistics}
\end{equation}
\end{widetext}
which we show in Fig.~\ref{Large deviation plots}(b) for three different couplings. 

To better understand these results, we return to the $s$-biased current in Fig.~\ref{Large deviation plots}(a), where we see that it diverges at the nonanalytic point of the scaled cumulant generating function, and those nonanalytic points are directly related to the convergence points found in Fig.~\ref{Fig2}(b,c). Thus, for large currents, $J\gg\bar \kappa$, the saddle-point is given by the convergence point as $s_\mathrm{sp}(J)\simeq e^{s_c}-1 \simeq s_c$ for $s_c\ll 1$, allowing us to approximate the tail of the large-deviation function by the straight line,
\begin{equation}
    \frac{\ln[P(J,t)]}{t} \simeq \Theta(s_c) -s_c J, \quad J\gg\bar \kappa,
\label{eq:LDFapprox}
\end{equation}
which we show in Fig.~\ref{Large deviation plots}(b) with dashes. The slope of the line is given by $s_c$, which becomes smaller as the coupling approaches its critical value, and $s_c$ goes to zero. Thus, we see that the extracted convergence points determine the tails of the large-deviation statistics.

The same phenomenon can be observed for the large-deviation statistics in the superradiant phase shown in Fig.~\ref{Large deviation plots}(c). In that case,  we rely on a numerical solution of the saddle-point equation. Still, we again see that the tails of the distribution are determined by the convergence points as shown by dashed lines. Moreover, in the superradiant phase, the bulk of the distribution becomes increasingly Poissonian, as the number of atoms is increased according to Eq.~(\ref{eq:scaledCGF}). Thus, for the sake of comparison, we show in Fig.~\ref{Large deviation plots}(c) the large-deviation statistics for  $N=100$ and $N=500$ atoms together with Poisson distributions that are represented by dotted lines. 

\section{Conclusions and outlook}
\label{sec:conclusion}

We have investigated the superradiant phase transition in the open Dicke model using our Lee-Yang theory of phase transitions. Specifically, we have shown how the dominant pole of the factorial cumulant generating function can be extracted from the high factorial cumulants of the photon emission statistics obtained during a finite observation time. As such, our method makes it possible to detect the superradiant phase transition, which occurs in the limit of long times, from measurements of the photon emission statistics, which are limited to finite durations. We have also shown how the convergence point of the dominant pole in the long-time limit determines the tails of the large-deviation statistics. Our method is not restricted to the Dicke model, and it can be applied to other systems that exhibit phase transitions, such as the  Rabi model~\cite{sahel13, Hwang:2015,Hwang:2018,zheng24,Ge:2024} or the Lipkin-Meshkov-Glick model~\cite{Lipkin:1965,Meshkov:1965,Glick:1965,Kopylov:2017}, both in theory and experiments.

We have seen that the system transits directly from the normal phase to the superradiant phase at a specific critical coupling, if the emission rate from the cavity is small. However, for larger decay rates, the critical coupling splits into two different values, and a small gap develops between the two phases. Here, we have applied our Lee-Yang theory to each of the phases, where the system can be mapped onto two uncoupled quantum harmonic oscillators, and the problem 
can be solved. By contrast, we are not aware of a solution in the gap region, which would allow us to calculate the  factorial cumulants and thereby apply our Lee-Yang method. However, \revision{theoretically}, it may be possible to explore this region using a different diagonalization procedure~\cite{PhysRevA.87.043840}, higher-order operator-cumulant expansions beyond the mean-field approximation~\cite{Kirton2018}, or advanced numerical methods for systems of finite size. \revision{Also, measurements of the photon counting statistics for an open Dicke model may shed further light on the unexplored region between the two critical points using our Lee-Yang formalism.}  We leave these tasks as interesting open problems for future work.

\section{Acknowledgments}

We acknowledge support from Jane and Aatos Erkko Foundation, Research Council of Finland through the Finnish Centre of Excellence in Quantum Technology (352925), and the Japan Society for the Promotion of Science through an Invitational Fellowship for Research in Japan. F.~N.~is supported in part by Nippon Telegraph and Telephone Corporation (NTT) Research, the Japan Science and Technology Agency (JST) [via the Quantum Leap Flagship Program (Q-LEAP), and the Moonshot R\&D Grant Number JPMJMS2061], the Asian Office of Aerospace Research and Development (AOARD) (via Grant No.~FA2386-20-1-4069), and the Office of Naval Research Global (ONR) (via Grant No.~N62909-23-1-2074). N.~L.~is supported by the RIKEN Incentive Research Program and by MEXT KAKENHI Grant Numbers JP24H00816, JP24H00820.

\appendix
\section{Diagonalization parameters}
\label{AppA}
We here provide the parameters for the  diagonalization in Eqs.~\eqref{Diagonalization} and~\eqref{Eigenvalue decomposed Hamiltonian}. To this end, we introduce the 
dimensionless parameters, $\tilde \omega_{1,2} = \omega_{1,2}/\omega_c$, with similar meanings for other parameters with a tilde. 

In the normal phase, we have
\begin{equation}
\varepsilon_{1_\pm} = \sin(\gamma)\frac{1\mp\tilde\omega_1}{2\sqrt{\tilde \omega_1}},\qquad\varepsilon_{2_\pm} = \cos(\gamma)\frac{1\mp\tilde\omega_2}{2\sqrt{\tilde\omega_2}},
\end{equation}
with 
\begin{equation}
\tilde \omega_1 =\sqrt{\left(1 +\tilde \omega_a^2 \pm\sqrt{\left(\tilde \omega_a^2-1\right)^2+16 \tilde \lambda^2 \tilde \omega_a }\right)/2},
\end{equation}
and
\begin{equation}
\tilde \omega_2 =\sqrt{\left(1 +\tilde \omega_a^2 \mp\sqrt{\left(\tilde \omega_a^2-1\right)^2+16 \tilde \lambda^2 \tilde \omega_a }\right)/2},
\end{equation}
where the signs are plus for $\tilde \omega_a >1$ and minus for $\tilde \omega_a\leq 1$.  Furthermore, the parameter $\gamma$ is obtained from
\begin{equation}
\tan(2\gamma) = 4\tilde \lambda \frac{ \sqrt{\tilde \omega_a}}{\tilde \omega_a^2-1}.
\end{equation}

For the superradiant phase, the eigenfrequencies are
\begin{equation}
\tilde\omega_1 =\sqrt{\left(1+4 \tilde\lambda_\mathrm{aa}\tilde \Omega_a+\tilde\Omega_a^2\pm\sqrt{\alpha} \right)/2},
\end{equation}
and
\begin{equation}
\tilde\omega_2 =\sqrt{\left(1+4 \tilde\lambda_\mathrm{aa}\tilde \Omega_a+\tilde\Omega_a^2\mp\sqrt{\alpha} \right)/2},
\end{equation}
where the signs are determined in the same way as for the normal phase. We have also introduced the parameter
\begin{equation}
    \alpha =1 +8\left(2\tilde\lambda_\mathrm{ca}^2-\tilde \lambda_\mathrm{aa}\right)\tilde\Omega_a
    -2\left(1-8 \tilde\lambda_\mathrm{aa}^2\right)\tilde \Omega_a^2 +8\tilde \lambda_\mathrm{aa} \tilde \Omega_a^3+\tilde\Omega_a^4,
\end{equation}
and the parameter $\gamma$ is obtained from
\begin{equation}
    \cos(\gamma)=\sqrt{\frac{\tilde \Omega_a^2 }{2\sqrt{\alpha}}+\frac{2}{\sqrt{\alpha}}\tilde \lambda_\mathrm{aa}\tilde \Omega_a-\frac{1}{2}\left( 1+\frac{1}{\sqrt{\alpha}}\right)}.
\end{equation}


\begin{thebibliography}{67}%
	\makeatletter
	\providecommand \@ifxundefined [1]{%
		\@ifx{#1\undefined}
	}%
	\providecommand \@ifnum [1]{%
		\ifnum #1\expandafter \@firstoftwo
		\else \expandafter \@secondoftwo
		\fi
	}%
	\providecommand \@ifx [1]{%
		\ifx #1\expandafter \@firstoftwo
		\else \expandafter \@secondoftwo
		\fi
	}%
	\providecommand \natexlab [1]{#1}%
	\providecommand \enquote  [1]{``#1''}%
	\providecommand \bibnamefont  [1]{#1}%
	\providecommand \bibfnamefont [1]{#1}%
	\providecommand \citenamefont [1]{#1}%
	\providecommand \href@noop [0]{\@secondoftwo}%
	\providecommand \href [0]{\begingroup \@sanitize@url \@href}%
	\providecommand \@href[1]{\@@startlink{#1}\@@href}%
	\providecommand \@@href[1]{\endgroup#1\@@endlink}%
	\providecommand \@sanitize@url [0]{\catcode `\\12\catcode `\$12\catcode
		`\&12\catcode `\#12\catcode `\^12\catcode `\_12\catcode `\%12\relax}%
	\providecommand \@@startlink[1]{}%
	\providecommand \@@endlink[0]{}%
	\providecommand \url  [0]{\begingroup\@sanitize@url \@url }%
	\providecommand \@url [1]{\endgroup\@href {#1}{\urlprefix }}%
	\providecommand \urlprefix  [0]{URL }%
	\providecommand \Eprint [0]{\href }%
	\providecommand \doibase [0]{https://doi.org/}%
	\providecommand \selectlanguage [0]{\@gobble}%
	\providecommand \bibinfo  [0]{\@secondoftwo}%
	\providecommand \bibfield  [0]{\@secondoftwo}%
	\providecommand \translation [1]{[#1]}%
	\providecommand \BibitemOpen [0]{}%
	\providecommand \bibitemStop [0]{}%
	\providecommand \bibitemNoStop [0]{.\EOS\space}%
	\providecommand \EOS [0]{\spacefactor3000\relax}%
	\providecommand \BibitemShut  [1]{\csname bibitem#1\endcsname}%
	\let\auto@bib@innerbib\@empty
	\bibitem [{\citenamefont {Dicke}(1954)}]{dicke}%
	\BibitemOpen
	\bibfield  {author} {\bibinfo {author} {\bibfnamefont {R.~H.}\ \bibnamefont
			{Dicke}},\ }\bibfield  {title} {\bibinfo {title} {{Coherence in Spontaneous
				Radiation Processes}},\ }\href {https://doi.org/10.1103/PhysRev.93.99}
	{\bibfield  {journal} {\bibinfo  {journal} {Phys. Rev.}\ }\textbf {\bibinfo
			{volume} {93}},\ \bibinfo {pages} {99} (\bibinfo {year} {1954})}\BibitemShut
	{NoStop}%
	\bibitem [{\citenamefont {Hepp}\ and\ \citenamefont {Lieb}(1973)}]{Hepp1973}%
	\BibitemOpen
	\bibfield  {author} {\bibinfo {author} {\bibfnamefont {K.}~\bibnamefont
			{Hepp}}\ and\ \bibinfo {author} {\bibfnamefont {E.~H.}\ \bibnamefont
			{Lieb}},\ }\bibfield  {title} {\bibinfo {title} {{On the superradiant phase
				transition for molecules in a quantized radiation field: the Dicke maser
				model}},\ }\href {https://doi.org/10.1016/0003-4916(73)90039-0} {\bibfield
		{journal} {\bibinfo  {journal} {Ann. Phys.}\ }\textbf {\bibinfo {volume}
			{76}},\ \bibinfo {pages} {360} (\bibinfo {year} {1973})}\BibitemShut
	{NoStop}%
	\bibitem [{\citenamefont {Wang}\ and\ \citenamefont {Hioe}(1973)}]{Wang1973}%
	\BibitemOpen
	\bibfield  {author} {\bibinfo {author} {\bibfnamefont {Y.~K.}\ \bibnamefont
			{Wang}}\ and\ \bibinfo {author} {\bibfnamefont {F.~T.}\ \bibnamefont
			{Hioe}},\ }\bibfield  {title} {\bibinfo {title} {{Phase Transition in the
				Dicke Model of Superradiance}},\ }\href
	{https://doi.org/10.1103/physreva.7.831} {\bibfield  {journal} {\bibinfo
			{journal} {Phys. Rev. A}\ }\textbf {\bibinfo {volume} {7}},\ \bibinfo {pages}
		{831} (\bibinfo {year} {1973})}\BibitemShut {NoStop}%
	\bibitem [{\citenamefont {Kirton}\ \emph {et~al.}(2018)\citenamefont {Kirton},
		\citenamefont {Roses}, \citenamefont {Keeling},\ and\ \citenamefont
		{Dalla~Torre}}]{Kirton2018}%
	\BibitemOpen
	\bibfield  {author} {\bibinfo {author} {\bibfnamefont {P.}~\bibnamefont
			{Kirton}}, \bibinfo {author} {\bibfnamefont {M.~M.}\ \bibnamefont {Roses}},
		\bibinfo {author} {\bibfnamefont {J.}~\bibnamefont {Keeling}},\ and\ \bibinfo
		{author} {\bibfnamefont {E.~G.}\ \bibnamefont {Dalla~Torre}},\ }\bibfield
	{title} {\bibinfo {title} {{Introduction to the Dicke Model: From Equilibrium
				to Nonequilibrium, and Vice Versa}},\ }\href
	{https://doi.org/10.1002/qute.201800043} {\bibfield  {journal} {\bibinfo
			{journal} {Adv. Quant. Technol.}\ }\textbf {\bibinfo {volume} {2}},\ \bibinfo
		{pages} {1970013} (\bibinfo {year} {2019})}\BibitemShut {NoStop}%
	\bibitem [{\citenamefont {Lambert}\ \emph {et~al.}(2009)\citenamefont
		{Lambert}, \citenamefont {Chen}, \citenamefont {Johansson},\ and\
		\citenamefont {Nori}}]{lambert09}%
	\BibitemOpen
	\bibfield  {author} {\bibinfo {author} {\bibfnamefont {N.}~\bibnamefont
			{Lambert}}, \bibinfo {author} {\bibfnamefont {Y.-N.}\ \bibnamefont {Chen}},
		\bibinfo {author} {\bibfnamefont {R.}~\bibnamefont {Johansson}},\ and\
		\bibinfo {author} {\bibfnamefont {F.}~\bibnamefont {Nori}},\ }\bibfield
	{title} {\bibinfo {title} {Quantum chaos and critical behavior on a chip},\
	}\href {https://doi.org/10.1103/PhysRevB.80.165308} {\bibfield  {journal}
		{\bibinfo  {journal} {Phys. Rev. B}\ }\textbf {\bibinfo {volume} {80}},\
		\bibinfo {pages} {165308} (\bibinfo {year} {2009})}\BibitemShut {NoStop}%
	\bibitem [{\citenamefont {Nataf}\ and\ \citenamefont
		{Ciuti}(2010)}]{Nataf2010}%
	\BibitemOpen
	\bibfield  {author} {\bibinfo {author} {\bibfnamefont {P.}~\bibnamefont
			{Nataf}}\ and\ \bibinfo {author} {\bibfnamefont {C.}~\bibnamefont {Ciuti}},\
	}\bibfield  {title} {\bibinfo {title} {No-go theorem for superradiant quantum
			phase transitions in cavity {QED} and counter-example in circuit {QED}},\
	}\href {https://doi.org/10.1038/ncomms1069} {\bibfield  {journal} {\bibinfo
			{journal} {Nat. Commun.}\ }\textbf {\bibinfo {volume} {1}},\ \bibinfo {pages}
		{72} (\bibinfo {year} {2010})}\BibitemShut {NoStop}%
	\bibitem [{\citenamefont {Viehmann}\ \emph {et~al.}(2011)\citenamefont
		{Viehmann}, \citenamefont {von Delft},\ and\ \citenamefont
		{Marquardt}}]{oliver11}%
	\BibitemOpen
	\bibfield  {author} {\bibinfo {author} {\bibfnamefont {O.}~\bibnamefont
			{Viehmann}}, \bibinfo {author} {\bibfnamefont {J.}~\bibnamefont {von
				Delft}},\ and\ \bibinfo {author} {\bibfnamefont {F.}~\bibnamefont
			{Marquardt}},\ }\bibfield  {title} {\bibinfo {title} {{Superradiant Phase
				Transitions and the Standard Description of Circuit QED}},\ }\href
	{https://doi.org/10.1103/PhysRevLett.107.113602} {\bibfield  {journal}
		{\bibinfo  {journal} {Phys. Rev. Lett.}\ }\textbf {\bibinfo {volume} {107}},\
		\bibinfo {pages} {113602} (\bibinfo {year} {2011})}\BibitemShut {NoStop}%
	\bibitem [{\citenamefont {Wang}\ \emph {et~al.}(2014)\citenamefont {Wang},
		\citenamefont {Wu}, \citenamefont {Yang}, \citenamefont {Jin}, \citenamefont
		{Lambert},\ and\ \citenamefont {Nori}}]{Wang2014}%
	\BibitemOpen
	\bibfield  {author} {\bibinfo {author} {\bibfnamefont {T.-L.}\ \bibnamefont
			{Wang}}, \bibinfo {author} {\bibfnamefont {L.-N.}\ \bibnamefont {Wu}},
		\bibinfo {author} {\bibfnamefont {W.}~\bibnamefont {Yang}}, \bibinfo {author}
		{\bibfnamefont {G.-R.}\ \bibnamefont {Jin}}, \bibinfo {author} {\bibfnamefont
			{N.}~\bibnamefont {Lambert}},\ and\ \bibinfo {author} {\bibfnamefont
			{F.}~\bibnamefont {Nori}},\ }\bibfield  {title} {\bibinfo {title} {{Quantum
				Fisher information as a signature of the superradiant quantum phase
				transition}},\ }\href {https://doi.org/10.1088/1367-2630/16/6/063039}
	{\bibfield  {journal} {\bibinfo  {journal} {New J. Phys.}\ }\textbf {\bibinfo
			{volume} {16}},\ \bibinfo {pages} {063039} (\bibinfo {year}
		{2014})}\BibitemShut {NoStop}%
	\bibitem [{\citenamefont {Lambert}\ \emph {et~al.}(2016)\citenamefont
		{Lambert}, \citenamefont {Matsuzaki}, \citenamefont {Kakuyanagi},
		\citenamefont {Ishida}, \citenamefont {Saito},\ and\ \citenamefont
		{Nori}}]{lambert16}%
	\BibitemOpen
	\bibfield  {author} {\bibinfo {author} {\bibfnamefont {N.}~\bibnamefont
			{Lambert}}, \bibinfo {author} {\bibfnamefont {Y.}~\bibnamefont {Matsuzaki}},
		\bibinfo {author} {\bibfnamefont {K.}~\bibnamefont {Kakuyanagi}}, \bibinfo
		{author} {\bibfnamefont {N.}~\bibnamefont {Ishida}}, \bibinfo {author}
		{\bibfnamefont {S.}~\bibnamefont {Saito}},\ and\ \bibinfo {author}
		{\bibfnamefont {F.}~\bibnamefont {Nori}},\ }\bibfield  {title} {\bibinfo
		{title} {Superradiance with an ensemble of superconducting flux qubits},\
	}\href {https://doi.org/10.1103/PhysRevB.94.224510} {\bibfield  {journal}
		{\bibinfo  {journal} {Phys. Rev. B}\ }\textbf {\bibinfo {volume} {94}},\
		\bibinfo {pages} {224510} (\bibinfo {year} {2016})}\BibitemShut {NoStop}%
	\bibitem [{\citenamefont {Bamba}\ \emph {et~al.}(2016)\citenamefont {Bamba},
		\citenamefont {Inomata},\ and\ \citenamefont {Nakamura}}]{bamba16}%
	\BibitemOpen
	\bibfield  {author} {\bibinfo {author} {\bibfnamefont {M.}~\bibnamefont
			{Bamba}}, \bibinfo {author} {\bibfnamefont {K.}~\bibnamefont {Inomata}},\
		and\ \bibinfo {author} {\bibfnamefont {Y.}~\bibnamefont {Nakamura}},\
	}\bibfield  {title} {\bibinfo {title} {{Superradiant Phase Transition in a
				Superconducting Circuit in Thermal Equilibrium}},\ }\href
	{https://doi.org/10.1103/PhysRevLett.117.173601} {\bibfield  {journal}
		{\bibinfo  {journal} {Phys. Rev. Lett.}\ }\textbf {\bibinfo {volume} {117}},\
		\bibinfo {pages} {173601} (\bibinfo {year} {2016})}\BibitemShut {NoStop}%
	\bibitem [{\citenamefont {Minganti}\ \emph {et~al.}(2021)\citenamefont
		{Minganti}, \citenamefont {Arkhipov}, \citenamefont {Miranowicz},\ and\
		\citenamefont {Nori}}]{Minganti:2021}%
	\BibitemOpen
	\bibfield  {author} {\bibinfo {author} {\bibfnamefont {F.}~\bibnamefont
			{Minganti}}, \bibinfo {author} {\bibfnamefont {I.~I.}\ \bibnamefont
			{Arkhipov}}, \bibinfo {author} {\bibfnamefont {A.}~\bibnamefont
			{Miranowicz}},\ and\ \bibinfo {author} {\bibfnamefont {F.}~\bibnamefont
			{Nori}},\ }\bibfield  {title} {\bibinfo {title} {Continuous dissipative phase
			transitions with or without symmetry breaking},\ }\href
	{https://doi.org/10.1088/1367-2630/ac3db8} {\bibfield  {journal} {\bibinfo
			{journal} {New J. Phys.}\ }\textbf {\bibinfo {volume} {23}},\ \bibinfo
		{pages} {122001} (\bibinfo {year} {2021})}\BibitemShut {NoStop}%
	\bibitem [{\citenamefont {De~Bernardis}\ \emph {et~al.}(2018)\citenamefont
		{De~Bernardis}, \citenamefont {Pilar}, \citenamefont {Jaako}, \citenamefont
		{De~Liberato},\ and\ \citenamefont {Rabl}}]{rabl18}%
	\BibitemOpen
	\bibfield  {author} {\bibinfo {author} {\bibfnamefont {D.}~\bibnamefont
			{De~Bernardis}}, \bibinfo {author} {\bibfnamefont {P.}~\bibnamefont {Pilar}},
		\bibinfo {author} {\bibfnamefont {T.}~\bibnamefont {Jaako}}, \bibinfo
		{author} {\bibfnamefont {S.}~\bibnamefont {De~Liberato}},\ and\ \bibinfo
		{author} {\bibfnamefont {P.}~\bibnamefont {Rabl}},\ }\bibfield  {title}
	{\bibinfo {title} {Breakdown of gauge invariance in ultrastrong-coupling
			cavity {QED}},\ }\href {https://doi.org/10.1103/PhysRevA.98.053819}
	{\bibfield  {journal} {\bibinfo  {journal} {Phys. Rev. A}\ }\textbf {\bibinfo
			{volume} {98}},\ \bibinfo {pages} {053819} (\bibinfo {year}
		{2018})}\BibitemShut {NoStop}%
	\bibitem [{\citenamefont {Di~Stefano}\ \emph {et~al.}(2019)\citenamefont
		{Di~Stefano}, \citenamefont {Settineri}, \citenamefont {Macrì},
		\citenamefont {Garziano}, \citenamefont {Stassi}, \citenamefont {Savasta},\
		and\ \citenamefont {Nori}}]{DiStefano2019}%
	\BibitemOpen
	\bibfield  {author} {\bibinfo {author} {\bibfnamefont {O.}~\bibnamefont
			{Di~Stefano}}, \bibinfo {author} {\bibfnamefont {A.}~\bibnamefont
			{Settineri}}, \bibinfo {author} {\bibfnamefont {V.}~\bibnamefont {Macrì}},
		\bibinfo {author} {\bibfnamefont {L.}~\bibnamefont {Garziano}}, \bibinfo
		{author} {\bibfnamefont {R.}~\bibnamefont {Stassi}}, \bibinfo {author}
		{\bibfnamefont {S.}~\bibnamefont {Savasta}},\ and\ \bibinfo {author}
		{\bibfnamefont {F.}~\bibnamefont {Nori}},\ }\bibfield  {title} {\bibinfo
		{title} {Resolution of gauge ambiguities in ultrastrong-coupling cavity
			quantum electrodynamics},\ }\href {https://doi.org/10.1038/s41567-019-0534-4}
	{\bibfield  {journal} {\bibinfo  {journal} {Nat. Phys.}\ }\textbf {\bibinfo
			{volume} {15}},\ \bibinfo {pages} {803} (\bibinfo {year} {2019})}\BibitemShut
	{NoStop}%
	\bibitem [{\citenamefont {Ashhab}\ \emph {et~al.}(2019)\citenamefont {Ashhab},
		\citenamefont {Matsuzaki}, \citenamefont {Kakuyanagi}, \citenamefont {Saito},
		\citenamefont {Yoshihara}, \citenamefont {Fuse},\ and\ \citenamefont
		{Semba}}]{sahel}%
	\BibitemOpen
	\bibfield  {author} {\bibinfo {author} {\bibfnamefont {S.}~\bibnamefont
			{Ashhab}}, \bibinfo {author} {\bibfnamefont {Y.}~\bibnamefont {Matsuzaki}},
		\bibinfo {author} {\bibfnamefont {K.}~\bibnamefont {Kakuyanagi}}, \bibinfo
		{author} {\bibfnamefont {S.}~\bibnamefont {Saito}}, \bibinfo {author}
		{\bibfnamefont {F.}~\bibnamefont {Yoshihara}}, \bibinfo {author}
		{\bibfnamefont {T.}~\bibnamefont {Fuse}},\ and\ \bibinfo {author}
		{\bibfnamefont {K.}~\bibnamefont {Semba}},\ }\bibfield  {title} {\bibinfo
		{title} {Spectrum of the {D}icke model in a superconducting qubit-oscillator
			system},\ }\href {https://doi.org/10.1103/PhysRevA.99.063822} {\bibfield
		{journal} {\bibinfo  {journal} {Phys. Rev. A}\ }\textbf {\bibinfo {volume}
			{99}},\ \bibinfo {pages} {063822} (\bibinfo {year} {2019})}\BibitemShut
	{NoStop}%
	\bibitem [{\citenamefont {Genway}\ \emph {et~al.}(2014)\citenamefont {Genway},
		\citenamefont {Li}, \citenamefont {Ates}, \citenamefont {Lanyon},\ and\
		\citenamefont {Lesanovsky}}]{genway14}%
	\BibitemOpen
	\bibfield  {author} {\bibinfo {author} {\bibfnamefont {S.}~\bibnamefont
			{Genway}}, \bibinfo {author} {\bibfnamefont {W.}~\bibnamefont {Li}}, \bibinfo
		{author} {\bibfnamefont {C.}~\bibnamefont {Ates}}, \bibinfo {author}
		{\bibfnamefont {B.~P.}\ \bibnamefont {Lanyon}},\ and\ \bibinfo {author}
		{\bibfnamefont {I.}~\bibnamefont {Lesanovsky}},\ }\bibfield  {title}
	{\bibinfo {title} {{Generalized {D}icke Nonequilibrium Dynamics in Trapped
				Ions}},\ }\href {https://doi.org/10.1103/PhysRevLett.112.023603} {\bibfield
		{journal} {\bibinfo  {journal} {Phys. Rev. Lett.}\ }\textbf {\bibinfo
			{volume} {112}},\ \bibinfo {pages} {023603} (\bibinfo {year}
		{2014})}\BibitemShut {NoStop}%
	\bibitem [{\citenamefont {De~Liberato}\ and\ \citenamefont
		{Ciuti}(2013)}]{simone}%
	\BibitemOpen
	\bibfield  {author} {\bibinfo {author} {\bibfnamefont {S.}~\bibnamefont
			{De~Liberato}}\ and\ \bibinfo {author} {\bibfnamefont {C.}~\bibnamefont
			{Ciuti}},\ }\bibfield  {title} {\bibinfo {title} {{Quantum Phases of a
				Multimode Bosonic Field Coupled to Flat Electronic Bands}},\ }\href
	{https://doi.org/10.1103/PhysRevLett.110.133603} {\bibfield  {journal}
		{\bibinfo  {journal} {Phys. Rev. Lett.}\ }\textbf {\bibinfo {volume} {110}},\
		\bibinfo {pages} {133603} (\bibinfo {year} {2013})}\BibitemShut {NoStop}%
	\bibitem [{\citenamefont {Emary}\ and\ \citenamefont
		{Brandes}(2003{\natexlab{a}})}]{Emary2003}%
	\BibitemOpen
	\bibfield  {author} {\bibinfo {author} {\bibfnamefont {C.}~\bibnamefont
			{Emary}}\ and\ \bibinfo {author} {\bibfnamefont {T.}~\bibnamefont
			{Brandes}},\ }\bibfield  {title} {\bibinfo {title} {{Quantum Chaos Triggered
				by Precursors of a Quantum Phase Transition: The Dicke Model}},\ }\href
	{https://doi.org/10.1103/physrevlett.90.044101} {\bibfield  {journal}
		{\bibinfo  {journal} {Phys. Rev. Lett.}\ }\textbf {\bibinfo {volume} {90}},\
		\bibinfo {pages} {044101} (\bibinfo {year} {2003}{\natexlab{a}})}\BibitemShut
	{NoStop}%
	\bibitem [{\citenamefont {Emary}\ and\ \citenamefont
		{Brandes}(2003{\natexlab{b}})}]{PhysRevE.67.066203}%
	\BibitemOpen
	\bibfield  {author} {\bibinfo {author} {\bibfnamefont {C.}~\bibnamefont
			{Emary}}\ and\ \bibinfo {author} {\bibfnamefont {T.}~\bibnamefont
			{Brandes}},\ }\bibfield  {title} {\bibinfo {title} {Chaos and the quantum
			phase transition in the {Dicke} model},\ }\href
	{https://doi.org/10.1103/PhysRevE.67.066203} {\bibfield  {journal} {\bibinfo
			{journal} {Phys. Rev. E}\ }\textbf {\bibinfo {volume} {67}},\ \bibinfo
		{pages} {066203} (\bibinfo {year} {2003}{\natexlab{b}})}\BibitemShut
	{NoStop}%
	\bibitem [{\citenamefont {Lambert}\ \emph {et~al.}(2004)\citenamefont
		{Lambert}, \citenamefont {Emary},\ and\ \citenamefont
		{Brandes}}]{Lambert2004}%
	\BibitemOpen
	\bibfield  {author} {\bibinfo {author} {\bibfnamefont {N.}~\bibnamefont
			{Lambert}}, \bibinfo {author} {\bibfnamefont {C.}~\bibnamefont {Emary}},\
		and\ \bibinfo {author} {\bibfnamefont {T.}~\bibnamefont {Brandes}},\
	}\bibfield  {title} {\bibinfo {title} {{Entanglement and the Phase Transition
				in Single-Mode Superradiance}},\ }\href
	{https://doi.org/10.1103/physrevlett.92.073602} {\bibfield  {journal}
		{\bibinfo  {journal} {Phys. Rev. Lett.}\ }\textbf {\bibinfo {volume} {92}},\
		\bibinfo {pages} {073602} (\bibinfo {year} {2004})}\BibitemShut {NoStop}%
	\bibitem [{\citenamefont {Dimer}\ \emph {et~al.}(2007)\citenamefont {Dimer},
		\citenamefont {Estienne}, \citenamefont {Parkins},\ and\ \citenamefont
		{Carmichael}}]{carmichaelPRA}%
	\BibitemOpen
	\bibfield  {author} {\bibinfo {author} {\bibfnamefont {F.}~\bibnamefont
			{Dimer}}, \bibinfo {author} {\bibfnamefont {B.}~\bibnamefont {Estienne}},
		\bibinfo {author} {\bibfnamefont {A.~S.}\ \bibnamefont {Parkins}},\ and\
		\bibinfo {author} {\bibfnamefont {H.~J.}\ \bibnamefont {Carmichael}},\
	}\bibfield  {title} {\bibinfo {title} {Proposed realization of the
			{D}icke-model quantum phase transition in an optical cavity {QED} system},\
	}\href {https://doi.org/10.1103/PhysRevA.75.013804} {\bibfield  {journal}
		{\bibinfo  {journal} {Phys. Rev. A}\ }\textbf {\bibinfo {volume} {75}},\
		\bibinfo {pages} {013804} (\bibinfo {year} {2007})}\BibitemShut {NoStop}%
	\bibitem [{\citenamefont {Zhiqiang}\ \emph {et~al.}(2017)\citenamefont
		{Zhiqiang}, \citenamefont {Lee}, \citenamefont {Kumar}, \citenamefont
		{Arnold}, \citenamefont {Masson}, \citenamefont {Parkins},\ and\
		\citenamefont {Barrett}}]{Zhiqiang:17}%
	\BibitemOpen
	\bibfield  {author} {\bibinfo {author} {\bibfnamefont {Z.}~\bibnamefont
			{Zhiqiang}}, \bibinfo {author} {\bibfnamefont {C.~H.}\ \bibnamefont {Lee}},
		\bibinfo {author} {\bibfnamefont {R.}~\bibnamefont {Kumar}}, \bibinfo
		{author} {\bibfnamefont {K.~J.}\ \bibnamefont {Arnold}}, \bibinfo {author}
		{\bibfnamefont {S.~J.}\ \bibnamefont {Masson}}, \bibinfo {author}
		{\bibfnamefont {A.~S.}\ \bibnamefont {Parkins}},\ and\ \bibinfo {author}
		{\bibfnamefont {M.~D.}\ \bibnamefont {Barrett}},\ }\bibfield  {title}
	{\bibinfo {title} {Nonequilibrium phase transition in a spin-1 {D}icke
			model},\ }\href {https://doi.org/10.1364/OPTICA.4.000424} {\bibfield
		{journal} {\bibinfo  {journal} {Optica}\ }\textbf {\bibinfo {volume} {4}},\
		\bibinfo {pages} {424} (\bibinfo {year} {2017})}\BibitemShut {NoStop}%
	\bibitem [{\citenamefont {Baumann}\ \emph {et~al.}(2010)\citenamefont
		{Baumann}, \citenamefont {Guerlin}, \citenamefont {Brennecke},\ and\
		\citenamefont {Esslinger}}]{Baumann2010}%
	\BibitemOpen
	\bibfield  {author} {\bibinfo {author} {\bibfnamefont {K.}~\bibnamefont
			{Baumann}}, \bibinfo {author} {\bibfnamefont {C.}~\bibnamefont {Guerlin}},
		\bibinfo {author} {\bibfnamefont {F.}~\bibnamefont {Brennecke}},\ and\
		\bibinfo {author} {\bibfnamefont {T.}~\bibnamefont {Esslinger}},\ }\bibfield
	{title} {\bibinfo {title} {{D}icke quantum phase transition with a superfluid
			gas in an optical cavity},\ }\href {https://doi.org/10.1038/nature09009}
	{\bibfield  {journal} {\bibinfo  {journal} {Nature}\ }\textbf {\bibinfo
			{volume} {464}},\ \bibinfo {pages} {1301} (\bibinfo {year}
		{2010})}\BibitemShut {NoStop}%
	\bibitem [{\citenamefont {Nagy}\ \emph {et~al.}(2010)\citenamefont {Nagy},
		\citenamefont {K\'onya}, \citenamefont {Szirmai},\ and\ \citenamefont
		{Domokos}}]{nagyPRL}%
	\BibitemOpen
	\bibfield  {author} {\bibinfo {author} {\bibfnamefont {D.}~\bibnamefont
			{Nagy}}, \bibinfo {author} {\bibfnamefont {G.}~\bibnamefont {K\'onya}},
		\bibinfo {author} {\bibfnamefont {G.}~\bibnamefont {Szirmai}},\ and\ \bibinfo
		{author} {\bibfnamefont {P.}~\bibnamefont {Domokos}},\ }\bibfield  {title}
	{\bibinfo {title} {{Dicke-Model Phase Transition in the Quantum Motion of a
				{B}ose-{E}instein Condensate in an Optical Cavity}},\ }\href
	{https://doi.org/10.1103/PhysRevLett.104.130401} {\bibfield  {journal}
		{\bibinfo  {journal} {Phys. Rev. Lett.}\ }\textbf {\bibinfo {volume} {104}},\
		\bibinfo {pages} {130401} (\bibinfo {year} {2010})}\BibitemShut {NoStop}%
	\bibitem [{\citenamefont {Torre}\ \emph {et~al.}(2013)\citenamefont {Torre},
		\citenamefont {Diehl}, \citenamefont {Lukin}, \citenamefont {Sachdev},\ and\
		\citenamefont {Strack}}]{dallaPRA}%
	\BibitemOpen
	\bibfield  {author} {\bibinfo {author} {\bibfnamefont {E.~G.~D.}\
			\bibnamefont {Torre}}, \bibinfo {author} {\bibfnamefont {S.}~\bibnamefont
			{Diehl}}, \bibinfo {author} {\bibfnamefont {M.~D.}\ \bibnamefont {Lukin}},
		\bibinfo {author} {\bibfnamefont {S.}~\bibnamefont {Sachdev}},\ and\ \bibinfo
		{author} {\bibfnamefont {P.}~\bibnamefont {Strack}},\ }\bibfield  {title}
	{\bibinfo {title} {Keldysh approach for nonequilibrium phase transitions in
			quantum optics: {B}eyond the {D}icke model in optical cavities},\ }\href
	{https://doi.org/10.1103/PhysRevA.87.023831} {\bibfield  {journal} {\bibinfo
			{journal} {Phys. Rev. A}\ }\textbf {\bibinfo {volume} {87}},\ \bibinfo
		{pages} {023831} (\bibinfo {year} {2013})}\BibitemShut {NoStop}%
			\bibitem [{\citenamefont {Carollo}\ \emph {et~al.}(2021)\citenamefont {Carollo},
			\citenamefont {Diehl},\ and\
			\citenamefont {Lesanovski}}]{Carollo:2021}%
		\BibitemOpen
		\bibfield  {author} {\bibinfo {author} {\bibfnamefont {F.}~\bibnamefont
				{Carollo}},\ and\ \bibinfo
			{author} {\bibfnamefont {I.}~\bibnamefont {Lesanovsky}},\ }\bibfield  {title}
		{\bibinfo {title} {{Exactness of Mean-Field Equations for Open Dicke Models with an Application to Pattern Retrieval Dynamics}},\ }\href
		{https://doi.org/10.1103/PhysRevLett.126.230601} {\bibfield  {journal}
			{\bibinfo  {journal} {Phys. Rev. Lett.}\ }\textbf {\bibinfo {volume} {126}},\
			\bibinfo {pages} {230601} (\bibinfo {year} {2021})}\BibitemShut {NoStop}%
					\bibitem [{\citenamefont {Boneberg}\ \emph {et~al.}(2022)\citenamefont {Carollo},
				\citenamefont {Diehl},\ and\
				\citenamefont {Lesanovski}}]{Boneberg:2022}%
			\BibitemOpen
			\bibfield  {author} {\bibinfo {author} {\bibfnamefont {M.}~\bibnamefont
					{Boneberg}}, \bibinfo {author} {\bibfnamefont {I.}\ \bibnamefont {Lesanovsky}},\ and\ \bibinfo
				{author} {\bibfnamefont {F.}~\bibnamefont {Carollo}},\ }\bibfield  {title}
			{\bibinfo {title} {{Quantum fluctuations and correlations in open quantum Dicke models}},\ }\href
			{https://doi.org/10.1103/PhysRevA.106.012212} {\bibfield  {journal}
				{\bibinfo  {journal} {Phys. Rev. A}\ }\textbf {\bibinfo {volume} {106}},\
				\bibinfo {pages} {012212} (\bibinfo {year} {2022})}\BibitemShut {NoStop}%
	\bibitem [{\citenamefont {Chen}\ \emph {et~al.}(2024)\citenamefont {Chen},
		\citenamefont {Qiu}, \citenamefont {Miranowicz}, \citenamefont {Lambert},
		\citenamefont {Qin}, \citenamefont {Stassi}, \citenamefont {Xia},
		\citenamefont {Zheng},\ and\ \citenamefont {Nori}}]{Chen2024}%
	\BibitemOpen
	\bibfield  {author} {\bibinfo {author} {\bibfnamefont {Y.-H.}\ \bibnamefont
			{Chen}}, \bibinfo {author} {\bibfnamefont {Y.}~\bibnamefont {Qiu}}, \bibinfo
		{author} {\bibfnamefont {A.}~\bibnamefont {Miranowicz}}, \bibinfo {author}
		{\bibfnamefont {N.}~\bibnamefont {Lambert}}, \bibinfo {author} {\bibfnamefont
			{W.}~\bibnamefont {Qin}}, \bibinfo {author} {\bibfnamefont {R.}~\bibnamefont
			{Stassi}}, \bibinfo {author} {\bibfnamefont {Y.}~\bibnamefont {Xia}},
		\bibinfo {author} {\bibfnamefont {S.-B.}\ \bibnamefont {Zheng}},\ and\
		\bibinfo {author} {\bibfnamefont {F.}~\bibnamefont {Nori}},\ }\bibfield
	{title} {\bibinfo {title} {{Sudden change of the photon output field marks
				phase transitions in the quantum Rabi model}},\ }\href
	{https://doi.org/10.1038/s42005-023-01457-w} {\bibfield  {journal} {\bibinfo
			{journal} {Commun. Phys.}\ }\textbf {\bibinfo {volume} {7}},\ \bibinfo
		{pages} {5} (\bibinfo {year} {2024})}\BibitemShut {NoStop}%
	\bibitem [{\citenamefont {Yang}\ and\ \citenamefont {Lee}(1952)}]{Yang:1952}%
	\BibitemOpen
	\bibfield  {author} {\bibinfo {author} {\bibfnamefont {C.~N.}\ \bibnamefont
			{Yang}}\ and\ \bibinfo {author} {\bibfnamefont {T.~D.}\ \bibnamefont {Lee}},\
	}\bibfield  {title} {\bibinfo {title} {{Statistical Theory of Equations of
				State and Phase Transitions. I. Theory of Condensation}},\ }\href
	{https://doi.org/10.1103/PhysRev.87.404} {\bibfield  {journal} {\bibinfo
			{journal} {Phys. Rev.}\ }\textbf {\bibinfo {volume} {87}},\ \bibinfo {pages}
		{404} (\bibinfo {year} {1952})}\BibitemShut {NoStop}%
	\bibitem [{\citenamefont {Lee}\ and\ \citenamefont {Yang}(1952)}]{Lee:1952}%
	\BibitemOpen
	\bibfield  {author} {\bibinfo {author} {\bibfnamefont {T.~D.}\ \bibnamefont
			{Lee}}\ and\ \bibinfo {author} {\bibfnamefont {C.~N.}\ \bibnamefont {Yang}},\
	}\bibfield  {title} {\bibinfo {title} {{Statistical Theory of Equations of
				State and Phase Transitions. II. Lattice Gas and Ising Model}},\ }\href
	{https://doi.org/10.1103/PhysRev.87.410} {\bibfield  {journal} {\bibinfo
			{journal} {Phys. Rev.}\ }\textbf {\bibinfo {volume} {87}},\ \bibinfo {pages}
		{410} (\bibinfo {year} {1952})}\BibitemShut {NoStop}%
	\bibitem [{\citenamefont {Blythe}\ and\ \citenamefont
		{Evans}(2003)}]{Blythe2003}%
	\BibitemOpen
	\bibfield  {author} {\bibinfo {author} {\bibfnamefont {R.~A.}\ \bibnamefont
			{Blythe}}\ and\ \bibinfo {author} {\bibfnamefont {M.~R.}\ \bibnamefont
			{Evans}},\ }\bibfield  {title} {\bibinfo {title} {The {L}ee-{Y}ang theory of
			equilibrium and nonequilibrium phase transitions},\ }\href
	{https://doi.org/10.1590/S0103-97332003000300008} {\bibfield  {journal}
		{\bibinfo  {journal} {Braz. J. Phys.}\ }\textbf {\bibinfo {volume} {33}},\
		\bibinfo {pages} {464} (\bibinfo {year} {2003})}\BibitemShut {NoStop}%
	\bibitem [{\citenamefont {Bena}\ \emph {et~al.}(2005)\citenamefont {Bena},
		\citenamefont {Droz},\ and\ \citenamefont {Lipowski}}]{Bena2005}%
	\BibitemOpen
	\bibfield  {author} {\bibinfo {author} {\bibfnamefont {I.}~\bibnamefont
			{Bena}}, \bibinfo {author} {\bibfnamefont {M.}~\bibnamefont {Droz}},\ and\
		\bibinfo {author} {\bibfnamefont {A.}~\bibnamefont {Lipowski}},\ }\bibfield
	{title} {\bibinfo {title} {Statistical mechanics of equilibrium and
			nonequilibrium phase transitions: The {Y}ang-{L}ee formalism},\ }\href
	{https://doi.org/10.1142/S0217979205032759} {\bibfield  {journal} {\bibinfo
			{journal} {Int. J. Mod. Phys. B}\ }\textbf {\bibinfo {volume} {19}},\
		\bibinfo {pages} {4269} (\bibinfo {year} {2005})}\BibitemShut {NoStop}%
	\bibitem [{\citenamefont {Blythe}\ and\ \citenamefont
		{Evans}(2002)}]{Blythe2002}%
	\BibitemOpen
	\bibfield  {author} {\bibinfo {author} {\bibfnamefont {R.~A.}\ \bibnamefont
			{Blythe}}\ and\ \bibinfo {author} {\bibfnamefont {M.~R.}\ \bibnamefont
			{Evans}},\ }\bibfield  {title} {\bibinfo {title} {{Lee-Yang Zeros and Phase
				Transitions in Nonequilibrium Steady States}},\ }\href
	{https://doi.org/10.1103/PhysRevLett.89.080601} {\bibfield  {journal}
		{\bibinfo  {journal} {Phys. Rev. Lett.}\ }\textbf {\bibinfo {volume} {89}},\
		\bibinfo {pages} {080601} (\bibinfo {year} {2002})}\BibitemShut {NoStop}%
	\bibitem [{\citenamefont {Flindt}\ and\ \citenamefont
		{Garrahan}(2013)}]{Flindt2013}%
	\BibitemOpen
	\bibfield  {author} {\bibinfo {author} {\bibfnamefont {C.}~\bibnamefont
			{Flindt}}\ and\ \bibinfo {author} {\bibfnamefont {J.~P.}\ \bibnamefont
			{Garrahan}},\ }\bibfield  {title} {\bibinfo {title} {Trajectory {P}hase
			{T}ransitions, {L}ee-{Y}ang {Z}eros, and {H}igh-{O}rder {C}umulants in {F}ull
			{C}ounting {S}tatistics},\ }\href
	{https://doi.org/10.1103/PhysRevLett.110.050601} {\bibfield  {journal}
		{\bibinfo  {journal} {Phys. Rev. Lett.}\ }\textbf {\bibinfo {volume} {110}},\
		\bibinfo {pages} {050601} (\bibinfo {year} {2013})}\BibitemShut {NoStop}%
	\bibitem [{\citenamefont {Heyl}\ \emph {et~al.}(2013)\citenamefont {Heyl},
		\citenamefont {Polkovnikov},\ and\ \citenamefont {Kehrein}}]{Heyl:2013}%
	\BibitemOpen
	\bibfield  {author} {\bibinfo {author} {\bibfnamefont {M.}~\bibnamefont
			{Heyl}}, \bibinfo {author} {\bibfnamefont {A.}~\bibnamefont {Polkovnikov}},\
		and\ \bibinfo {author} {\bibfnamefont {S.}~\bibnamefont {Kehrein}},\
	}\bibfield  {title} {\bibinfo {title} {{Dynamical Quantum Phase Transitions
				in the Transverse-Field Ising Model}},\ }\href
	{https://doi.org/10.1103/PhysRevLett.110.135704} {\bibfield  {journal}
		{\bibinfo  {journal} {Phys. Rev. Lett.}\ }\textbf {\bibinfo {volume} {110}},\
		\bibinfo {pages} {135704} (\bibinfo {year} {2013})}\BibitemShut {NoStop}%
	\bibitem [{\citenamefont {Xu}\ \emph {et~al.}(2020)\citenamefont {Xu},
		\citenamefont {Sun}, \citenamefont {Liu}, \citenamefont {Zhang},
		\citenamefont {Li}, \citenamefont {Dong}, \citenamefont {Ren}, \citenamefont
		{Zhang}, \citenamefont {Nori}, \citenamefont {Zheng}, \citenamefont {Fan},\
		and\ \citenamefont {Wang}}]{Xu2020}%
	\BibitemOpen
	\bibfield  {author} {\bibinfo {author} {\bibfnamefont {K.}~\bibnamefont
			{Xu}}, \bibinfo {author} {\bibfnamefont {Z.-H.}\ \bibnamefont {Sun}},
		\bibinfo {author} {\bibfnamefont {W.}~\bibnamefont {Liu}}, \bibinfo {author}
		{\bibfnamefont {Y.-R.}\ \bibnamefont {Zhang}}, \bibinfo {author}
		{\bibfnamefont {H.}~\bibnamefont {Li}}, \bibinfo {author} {\bibfnamefont
			{H.}~\bibnamefont {Dong}}, \bibinfo {author} {\bibfnamefont {W.}~\bibnamefont
			{Ren}}, \bibinfo {author} {\bibfnamefont {P.}~\bibnamefont {Zhang}}, \bibinfo
		{author} {\bibfnamefont {F.}~\bibnamefont {Nori}}, \bibinfo {author}
		{\bibfnamefont {D.}~\bibnamefont {Zheng}}, \bibinfo {author} {\bibfnamefont
			{H.}~\bibnamefont {Fan}},\ and\ \bibinfo {author} {\bibfnamefont
			{H.}~\bibnamefont {Wang}},\ }\bibfield  {title} {\bibinfo {title} {Probing
			dynamical phase transitions with a superconducting quantum simulator},\
	}\href {https://doi.org/10.1126/sciadv.aba4935} {\bibfield  {journal}
		{\bibinfo  {journal} {Sci. Adv.}\ }\textbf {\bibinfo {volume} {6}},\ \bibinfo
		{pages} {eaba4935} (\bibinfo {year} {2020})}\BibitemShut {NoStop}%
	\bibitem [{\citenamefont {Peotta}\ \emph {et~al.}(2021)\citenamefont {Peotta},
		\citenamefont {Brange}, \citenamefont {Deger}, \citenamefont {Ojanen},\ and\
		\citenamefont {Flindt}}]{Peotta2021}%
	\BibitemOpen
	\bibfield  {author} {\bibinfo {author} {\bibfnamefont {S.}~\bibnamefont
			{Peotta}}, \bibinfo {author} {\bibfnamefont {F.}~\bibnamefont {Brange}},
		\bibinfo {author} {\bibfnamefont {A.}~\bibnamefont {Deger}}, \bibinfo
		{author} {\bibfnamefont {T.}~\bibnamefont {Ojanen}},\ and\ \bibinfo {author}
		{\bibfnamefont {C.}~\bibnamefont {Flindt}},\ }\bibfield  {title} {\bibinfo
		{title} {Determination of {D}ynamical {Q}uantum {P}hase {T}ransitions in
			{S}trongly {C}orrelated {M}any-{B}ody {S}ystems {U}sing {L}oschmidt
			{C}umulants},\ }\href {https://doi.org/10.1103/PhysRevX.11.041018} {\bibfield
		{journal} {\bibinfo  {journal} {Phys. Rev. X}\ }\textbf {\bibinfo {volume}
			{11}},\ \bibinfo {pages} {041018} (\bibinfo {year} {2021})}\BibitemShut
	{NoStop}%
	\bibitem [{\citenamefont {Brange}\ \emph {et~al.}(2022)\citenamefont {Brange},
		\citenamefont {Peotta}, \citenamefont {Flindt},\ and\ \citenamefont
		{Ojanen}}]{Brange:2022}%
	\BibitemOpen
	\bibfield  {author} {\bibinfo {author} {\bibfnamefont {F.}~\bibnamefont
			{Brange}}, \bibinfo {author} {\bibfnamefont {S.}~\bibnamefont {Peotta}},
		\bibinfo {author} {\bibfnamefont {C.}~\bibnamefont {Flindt}},\ and\ \bibinfo
		{author} {\bibfnamefont {T.}~\bibnamefont {Ojanen}},\ }\bibfield  {title}
	{\bibinfo {title} {Dynamical quantum phase transitions in strongly correlated
			two-dimensional spin lattices following a quench},\ }\href
	{https://doi.org/10.1103/PhysRevResearch.4.033032} {\bibfield  {journal}
		{\bibinfo  {journal} {Phys. Rev. Res.}\ }\textbf {\bibinfo {volume} {4}},\
		\bibinfo {pages} {033032} (\bibinfo {year} {2022})}\BibitemShut {NoStop}%
	\bibitem [{\citenamefont {Kist}\ \emph {et~al.}(2021)\citenamefont {Kist},
		\citenamefont {Lado},\ and\ \citenamefont {Flindt}}]{Kist2021}%
	\BibitemOpen
	\bibfield  {author} {\bibinfo {author} {\bibfnamefont {T.}~\bibnamefont
			{Kist}}, \bibinfo {author} {\bibfnamefont {J.~L.}\ \bibnamefont {Lado}},\
		and\ \bibinfo {author} {\bibfnamefont {C.}~\bibnamefont {Flindt}},\
	}\bibfield  {title} {\bibinfo {title} {{Lee-Yang theory of criticality in
				interacting quantum many-body systems}},\ }\href
	{https://doi.org/10.1103/PhysRevResearch.3.033206} {\bibfield  {journal}
		{\bibinfo  {journal} {Phys. Rev. Res.}\ }\textbf {\bibinfo {volume} {3}},\
		\bibinfo {pages} {033206} (\bibinfo {year} {2021})}\BibitemShut {NoStop}%
	\bibitem [{\citenamefont {Vecsei}\ \emph {et~al.}(2022)\citenamefont {Vecsei},
		\citenamefont {Lado},\ and\ \citenamefont {Flindt}}]{Vecsei2022}%
	\BibitemOpen
	\bibfield  {author} {\bibinfo {author} {\bibfnamefont {P.~M.}\ \bibnamefont
			{Vecsei}}, \bibinfo {author} {\bibfnamefont {J.~L.}\ \bibnamefont {Lado}},\
		and\ \bibinfo {author} {\bibfnamefont {C.}~\bibnamefont {Flindt}},\
	}\bibfield  {title} {\bibinfo {title} {{Lee-Yang theory of the
				two-dimensional quantum Ising model}},\ }\href
	{https://doi.org/10.1103/PhysRevB.106.054402} {\bibfield  {journal} {\bibinfo
			{journal} {Phys. Rev. B}\ }\textbf {\bibinfo {volume} {106}},\ \bibinfo
		{pages} {054402} (\bibinfo {year} {2022})}\BibitemShut {NoStop}%
	\bibitem [{\citenamefont {Vecsei}\ \emph {et~al.}(2023)\citenamefont {Vecsei},
		\citenamefont {Flindt},\ and\ \citenamefont {Lado}}]{Vecsei:2023}%
	\BibitemOpen
	\bibfield  {author} {\bibinfo {author} {\bibfnamefont {P.~M.}\ \bibnamefont
			{Vecsei}}, \bibinfo {author} {\bibfnamefont {C.}~\bibnamefont {Flindt}},\
		and\ \bibinfo {author} {\bibfnamefont {J.~L.}\ \bibnamefont {Lado}},\
	}\bibfield  {title} {\bibinfo {title} {{Lee-Yang theory of quantum phase
				transitions with neural network quantum states}},\ }\href
	{https://doi.org/10.1103/PhysRevResearch.5.033116} {\bibfield  {journal}
		{\bibinfo  {journal} {Phys. Rev. Res.}\ }\textbf {\bibinfo {volume} {5}},\
		\bibinfo {pages} {033116} (\bibinfo {year} {2023})}\BibitemShut {NoStop}%
	\bibitem [{\citenamefont {Binek}(1998)}]{Binek1998}%
	\BibitemOpen
	\bibfield  {author} {\bibinfo {author} {\bibfnamefont {C.}~\bibnamefont
			{Binek}},\ }\bibfield  {title} {\bibinfo {title} {{Density of Zeros on the
				Lee-Yang Circle Obtained from Magnetization Data of a Two-Dimensional Ising
				Ferromagnet}},\ }\href {https://doi.org/10.1103/PhysRevLett.81.5644}
	{\bibfield  {journal} {\bibinfo  {journal} {Phys. Rev. Lett.}\ }\textbf
		{\bibinfo {volume} {81}},\ \bibinfo {pages} {5644} (\bibinfo {year}
		{1998})}\BibitemShut {NoStop}%
	\bibitem [{\citenamefont {Peng}\ \emph {et~al.}(2015)\citenamefont {Peng},
		\citenamefont {Zhou}, \citenamefont {Wei}, \citenamefont {Cui}, \citenamefont
		{Du},\ and\ \citenamefont {Liu}}]{Peng2015}%
	\BibitemOpen
	\bibfield  {author} {\bibinfo {author} {\bibfnamefont {X.}~\bibnamefont
			{Peng}}, \bibinfo {author} {\bibfnamefont {H.}~\bibnamefont {Zhou}}, \bibinfo
		{author} {\bibfnamefont {B.-B.}\ \bibnamefont {Wei}}, \bibinfo {author}
		{\bibfnamefont {J.}~\bibnamefont {Cui}}, \bibinfo {author} {\bibfnamefont
			{J.}~\bibnamefont {Du}},\ and\ \bibinfo {author} {\bibfnamefont {R.-B.}\
			\bibnamefont {Liu}},\ }\bibfield  {title} {\bibinfo {title} {{Experimental
				Observation of Lee-Yang Zeros}},\ }\href
	{https://doi.org/10.1103/PhysRevLett.114.010601} {\bibfield  {journal}
		{\bibinfo  {journal} {Phys. Rev. Lett.}\ }\textbf {\bibinfo {volume} {114}},\
		\bibinfo {pages} {010601} (\bibinfo {year} {2015})}\BibitemShut {NoStop}%
	\bibitem [{\citenamefont {Brandner}\ \emph {et~al.}(2017)\citenamefont
		{Brandner}, \citenamefont {Maisi}, \citenamefont {Pekola}, \citenamefont
		{Garrahan},\ and\ \citenamefont {Flindt}}]{Brandner2017}%
	\BibitemOpen
	\bibfield  {author} {\bibinfo {author} {\bibfnamefont {K.}~\bibnamefont
			{Brandner}}, \bibinfo {author} {\bibfnamefont {V.~F.}\ \bibnamefont {Maisi}},
		\bibinfo {author} {\bibfnamefont {J.~P.}\ \bibnamefont {Pekola}}, \bibinfo
		{author} {\bibfnamefont {J.~P.}\ \bibnamefont {Garrahan}},\ and\ \bibinfo
		{author} {\bibfnamefont {C.}~\bibnamefont {Flindt}},\ }\bibfield  {title}
	{\bibinfo {title} {{Experimental Determination of Dynamical Lee-Yang
				Zeros}},\ }\href {https://doi.org/10.1103/physrevlett.118.180601} {\bibfield
		{journal} {\bibinfo  {journal} {Phys. Rev. Lett.}\ }\textbf {\bibinfo
			{volume} {118}},\ \bibinfo {pages} {180601} (\bibinfo {year}
		{2017})}\BibitemShut {NoStop}%
	\bibitem [{\citenamefont {Gao}\ \emph {et~al.}(2024)\citenamefont {Gao},
		\citenamefont {Wang}, \citenamefont {Xiao}, \citenamefont {Nakagawa},
		\citenamefont {Matsumoto}, \citenamefont {Qu}, \citenamefont {Lin},
		\citenamefont {Ueda},\ and\ \citenamefont {Xue}}]{gao:2024}%
	\BibitemOpen
	\bibfield  {author} {\bibinfo {author} {\bibfnamefont {H.}~\bibnamefont
			{Gao}}, \bibinfo {author} {\bibfnamefont {K.}~\bibnamefont {Wang}}, \bibinfo
		{author} {\bibfnamefont {L.}~\bibnamefont {Xiao}}, \bibinfo {author}
		{\bibfnamefont {M.}~\bibnamefont {Nakagawa}}, \bibinfo {author}
		{\bibfnamefont {N.}~\bibnamefont {Matsumoto}}, \bibinfo {author}
		{\bibfnamefont {D.}~\bibnamefont {Qu}}, \bibinfo {author} {\bibfnamefont
			{H.}~\bibnamefont {Lin}}, \bibinfo {author} {\bibfnamefont {M.}~\bibnamefont
			{Ueda}},\ and\ \bibinfo {author} {\bibfnamefont {P.}~\bibnamefont {Xue}},\
	}\bibfield  {title} {\bibinfo {title} {{Experimental Observation of the
				Yang-Lee Quantum Criticality in Open Quantum Systems}},\ }\href
	{https://doi.org/10.1103/PhysRevLett.132.176601} {\bibfield  {journal}
		{\bibinfo  {journal} {Phys. Rev. Lett.}\ }\textbf {\bibinfo {volume} {132}},\
		\bibinfo {pages} {176601} (\bibinfo {year} {2024})}\BibitemShut {NoStop}%
	\bibitem [{\citenamefont {Kopylov}\ \emph {et~al.}(2013)\citenamefont
		{Kopylov}, \citenamefont {Emary},\ and\ \citenamefont
		{Brandes}}]{PhysRevA.87.043840}%
	\BibitemOpen
	\bibfield  {author} {\bibinfo {author} {\bibfnamefont {W.}~\bibnamefont
			{Kopylov}}, \bibinfo {author} {\bibfnamefont {C.}~\bibnamefont {Emary}},\
		and\ \bibinfo {author} {\bibfnamefont {T.}~\bibnamefont {Brandes}},\
	}\bibfield  {title} {\bibinfo {title} {Counting statistics of the {Dicke}
			superradiance phase transition},\ }\href
	{https://doi.org/10.1103/PhysRevA.87.043840} {\bibfield  {journal} {\bibinfo
			{journal} {Phys. Rev. A}\ }\textbf {\bibinfo {volume} {87}},\ \bibinfo
		{pages} {043840} (\bibinfo {year} {2013})}\BibitemShut {NoStop}%
	\bibitem [{\citenamefont {Holstein}\ and\ \citenamefont
		{Primakoff}(1940)}]{PhysRev.58.1098}%
	\BibitemOpen
	\bibfield  {author} {\bibinfo {author} {\bibfnamefont {T.}~\bibnamefont
			{Holstein}}\ and\ \bibinfo {author} {\bibfnamefont {H.}~\bibnamefont
			{Primakoff}},\ }\bibfield  {title} {\bibinfo {title} {Field dependence of the
			intrinsic domain magnetization of a ferromagnet},\ }\href
	{https://doi.org/10.1103/PhysRev.58.1098} {\bibfield  {journal} {\bibinfo
			{journal} {Phys. Rev.}\ }\textbf {\bibinfo {volume} {58}},\ \bibinfo {pages}
		{1098} (\bibinfo {year} {1940})}\BibitemShut {NoStop}%
	\bibitem [{\citenamefont {Kapor}\ \emph {et~al.}(1991)\citenamefont {Kapor},
		\citenamefont {\ifmmode~\check{S}\else \v{S}\fi{}krinjar},\ and\
		\citenamefont {Stojanovi\ifmmode~\acute{c}\else
			\'{c}\fi{}}}]{PhysRevB.44.2227}%
	\BibitemOpen
	\bibfield  {author} {\bibinfo {author} {\bibfnamefont {D.~V.}\ \bibnamefont
			{Kapor}}, \bibinfo {author} {\bibfnamefont {M.~J.}\ \bibnamefont
			{\ifmmode~\check{S}\else \v{S}\fi{}krinjar}},\ and\ \bibinfo {author}
		{\bibfnamefont {S.~D.}\ \bibnamefont {Stojanovi\ifmmode~\acute{c}\else
				\'{c}\fi{}}},\ }\bibfield  {title} {\bibinfo {title} {Relation between
			spin-coherent states and boson-coherent states in the theory of magnetism},\
	}\href {https://doi.org/10.1103/PhysRevB.44.2227} {\bibfield  {journal}
		{\bibinfo  {journal} {Phys. Rev. B}\ }\textbf {\bibinfo {volume} {44}},\
		\bibinfo {pages} {2227} (\bibinfo {year} {1991})}\BibitemShut {NoStop}%
	\bibitem [{\citenamefont {Kambly}\ \emph {et~al.}(2011)\citenamefont {Kambly},
		\citenamefont {Flindt},\ and\ \citenamefont {B\"uttiker}}]{Kambly:2011}%
	\BibitemOpen
	\bibfield  {author} {\bibinfo {author} {\bibfnamefont {D.}~\bibnamefont
			{Kambly}}, \bibinfo {author} {\bibfnamefont {C.}~\bibnamefont {Flindt}},\
		and\ \bibinfo {author} {\bibfnamefont {M.}~\bibnamefont {B\"uttiker}},\
	}\bibfield  {title} {\bibinfo {title} {Factorial cumulants reveal
			interactions in counting statistics},\ }\href
	{https://doi.org/10.1103/PhysRevB.83.075432} {\bibfield  {journal} {\bibinfo
			{journal} {Phys. Rev. B}\ }\textbf {\bibinfo {volume} {83}},\ \bibinfo
		{pages} {075432} (\bibinfo {year} {2011})}\BibitemShut {NoStop}%
	\bibitem [{\citenamefont {Kambly}\ and\ \citenamefont
		{Flindt}(2013)}]{Kambly:2013}%
	\BibitemOpen
	\bibfield  {author} {\bibinfo {author} {\bibfnamefont {D.}~\bibnamefont
			{Kambly}}\ and\ \bibinfo {author} {\bibfnamefont {C.}~\bibnamefont
			{Flindt}},\ }\bibfield  {title} {\bibinfo {title} {Time-dependent factorial
			cumulants in interacting nano-scale systems},\ }\href
	{https://doi.org/10.1007/s10825-013-0464-9} {\bibfield  {journal} {\bibinfo
			{journal} {J. Comp. Elec.}\ }\textbf {\bibinfo {volume} {12}},\ \bibinfo
		{pages} {331} (\bibinfo {year} {2013})}\BibitemShut {NoStop}%
	\bibitem [{\citenamefont {Stegmann}\ \emph {et~al.}(2015)\citenamefont
		{Stegmann}, \citenamefont {Sothmann}, \citenamefont {Hucht},\ and\
		\citenamefont {K\"onig}}]{Stegmann:2015}%
	\BibitemOpen
	\bibfield  {author} {\bibinfo {author} {\bibfnamefont {P.}~\bibnamefont
			{Stegmann}}, \bibinfo {author} {\bibfnamefont {B.}~\bibnamefont {Sothmann}},
		\bibinfo {author} {\bibfnamefont {A.}~\bibnamefont {Hucht}},\ and\ \bibinfo
		{author} {\bibfnamefont {J.}~\bibnamefont {K\"onig}},\ }\bibfield  {title}
	{\bibinfo {title} {Detection of interactions via generalized factorial
			cumulants in systems in and out of equilibrium},\ }\href
	{https://doi.org/10.1103/PhysRevB.92.155413} {\bibfield  {journal} {\bibinfo
			{journal} {Phys. Rev. B}\ }\textbf {\bibinfo {volume} {92}},\ \bibinfo
		{pages} {155413} (\bibinfo {year} {2015})}\BibitemShut {NoStop}%
	\bibitem [{\citenamefont {König}\ and\ \citenamefont
		{Hucht}(2021)}]{Konig:2021}%
	\BibitemOpen
	\bibfield  {author} {\bibinfo {author} {\bibfnamefont {J.}~\bibnamefont
			{König}}\ and\ \bibinfo {author} {\bibfnamefont {A.}~\bibnamefont {Hucht}},\
	}\bibfield  {title} {\bibinfo {title} {{Newton series expansion of bosonic
				operator functions}},\ }\href {https://doi.org/10.21468/SciPostPhys.10.1.007}
	{\bibfield  {journal} {\bibinfo  {journal} {SciPost Phys.}\ }\textbf
		{\bibinfo {volume} {10}},\ \bibinfo {pages} {007} (\bibinfo {year}
		{2021})}\BibitemShut {NoStop}%
	\bibitem [{\citenamefont {Kleinherbers}\ \emph {et~al.}(2022)\citenamefont
		{Kleinherbers}, \citenamefont {Stegmann}, \citenamefont {Kurzmann},
		\citenamefont {Geller}, \citenamefont {Lorke},\ and\ \citenamefont
		{K\"onig}}]{Kleinherbers:2022}%
	\BibitemOpen
	\bibfield  {author} {\bibinfo {author} {\bibfnamefont {E.}~\bibnamefont
			{Kleinherbers}}, \bibinfo {author} {\bibfnamefont {P.}~\bibnamefont
			{Stegmann}}, \bibinfo {author} {\bibfnamefont {A.}~\bibnamefont {Kurzmann}},
		\bibinfo {author} {\bibfnamefont {M.}~\bibnamefont {Geller}}, \bibinfo
		{author} {\bibfnamefont {A.}~\bibnamefont {Lorke}},\ and\ \bibinfo {author}
		{\bibfnamefont {J.}~\bibnamefont {K\"onig}},\ }\bibfield  {title} {\bibinfo
		{title} {{Pushing the Limits in Real-Time Measurements of Quantum
				Dynamics}},\ }\href {https://doi.org/10.1103/PhysRevLett.128.087701}
	{\bibfield  {journal} {\bibinfo  {journal} {Phys. Rev. Lett.}\ }\textbf
		{\bibinfo {volume} {128}},\ \bibinfo {pages} {087701} (\bibinfo {year}
		{2022})}\BibitemShut {NoStop}%
	\bibitem [{\citenamefont {Plenio}\ and\ \citenamefont
		{Knight}(1998)}]{Plenio:1998}%
	\BibitemOpen
	\bibfield  {author} {\bibinfo {author} {\bibfnamefont {M.~B.}\ \bibnamefont
			{Plenio}}\ and\ \bibinfo {author} {\bibfnamefont {P.~L.}\ \bibnamefont
			{Knight}},\ }\bibfield  {title} {\bibinfo {title} {The quantum-jump approach
			to dissipative dynamics in quantum optics},\ }\href
	{https://doi.org/10.1103/RevModPhys.70.101} {\bibfield  {journal} {\bibinfo
			{journal} {Rev. Mod. Phys.}\ }\textbf {\bibinfo {volume} {70}},\ \bibinfo
		{pages} {101} (\bibinfo {year} {1998})}\BibitemShut {NoStop}%
	\bibitem [{\citenamefont {Brange}\ \emph {et~al.}(2019)\citenamefont {Brange},
		\citenamefont {Menczel},\ and\ \citenamefont {Flindt}}]{PhysRevB.99.085418}%
	\BibitemOpen
	\bibfield  {author} {\bibinfo {author} {\bibfnamefont {F.}~\bibnamefont
			{Brange}}, \bibinfo {author} {\bibfnamefont {P.}~\bibnamefont {Menczel}},\
		and\ \bibinfo {author} {\bibfnamefont {C.}~\bibnamefont {Flindt}},\
	}\bibfield  {title} {\bibinfo {title} {Photon counting statistics of a
			microwave cavity},\ }\href {https://doi.org/10.1103/PhysRevB.99.085418}
	{\bibfield  {journal} {\bibinfo  {journal} {Phys. Rev. B}\ }\textbf {\bibinfo
			{volume} {99}},\ \bibinfo {pages} {085418} (\bibinfo {year}
		{2019})}\BibitemShut {NoStop}%
	\bibitem [{\citenamefont {Deger}\ \emph {et~al.}(2018)\citenamefont {Deger},
		\citenamefont {Brandner},\ and\ \citenamefont {Flindt}}]{Deger2018}%
	\BibitemOpen
	\bibfield  {author} {\bibinfo {author} {\bibfnamefont {A.}~\bibnamefont
			{Deger}}, \bibinfo {author} {\bibfnamefont {K.}~\bibnamefont {Brandner}},\
		and\ \bibinfo {author} {\bibfnamefont {C.}~\bibnamefont {Flindt}},\
	}\bibfield  {title} {\bibinfo {title} {Lee-{Y}ang zeros and large-deviation
			statistics of a molecular zipper},\ }\href
	{https://doi.org/10.1103/physreve.97.012115} {\bibfield  {journal} {\bibinfo
			{journal} {Phys. Rev. E}\ }\textbf {\bibinfo {volume} {97}},\ \bibinfo
		{pages} {012115} (\bibinfo {year} {2018})}\BibitemShut {NoStop}%
	\bibitem [{\citenamefont {Brange}\ \emph {et~al.}(2023)\citenamefont {Brange},
		\citenamefont {Pyh\"aranta}, \citenamefont {Heinonen}, \citenamefont
		{Brandner},\ and\ \citenamefont {Flindt}}]{Brange:2023}%
	\BibitemOpen
	\bibfield  {author} {\bibinfo {author} {\bibfnamefont {F.}~\bibnamefont
			{Brange}}, \bibinfo {author} {\bibfnamefont {T.}~\bibnamefont {Pyh\"aranta}},
		\bibinfo {author} {\bibfnamefont {E.}~\bibnamefont {Heinonen}}, \bibinfo
		{author} {\bibfnamefont {K.}~\bibnamefont {Brandner}},\ and\ \bibinfo
		{author} {\bibfnamefont {C.}~\bibnamefont {Flindt}},\ }\bibfield  {title}
	{\bibinfo {title} {{Lee-Yang theory of Bose-Einstein condensation}},\ }\href
	{https://doi.org/10.1103/PhysRevA.107.033324} {\bibfield  {journal} {\bibinfo
			{journal} {Phys. Rev. A}\ }\textbf {\bibinfo {volume} {107}},\ \bibinfo
		{pages} {033324} (\bibinfo {year} {2023})}\BibitemShut {NoStop}%
	\bibitem [{\citenamefont {Wei}\ and\ \citenamefont {Liu}(2012)}]{Wei:2012}%
	\BibitemOpen
	\bibfield  {author} {\bibinfo {author} {\bibfnamefont {B.-B.}\ \bibnamefont
			{Wei}}\ and\ \bibinfo {author} {\bibfnamefont {R.-B.}\ \bibnamefont {Liu}},\
	}\bibfield  {title} {\bibinfo {title} {{Lee-Yang Zeros and Critical Times in
				Decoherence of a Probe Spin Coupled to a Bath}},\ }\href
	{https://doi.org/10.1103/PhysRevLett.109.185701} {\bibfield  {journal}
		{\bibinfo  {journal} {Phys. Rev. Lett.}\ }\textbf {\bibinfo {volume} {109}},\
		\bibinfo {pages} {185701} (\bibinfo {year} {2012})}\BibitemShut {NoStop}%
	\bibitem [{\citenamefont {Wei}\ \emph {et~al.}(2014)\citenamefont {Wei},
		\citenamefont {Chen}, \citenamefont {Po},\ and\ \citenamefont
		{Liu}}]{Wei:2014}%
	\BibitemOpen
	\bibfield  {author} {\bibinfo {author} {\bibfnamefont {B.-B.}\ \bibnamefont
			{Wei}}, \bibinfo {author} {\bibfnamefont {S.-W.}\ \bibnamefont {Chen}},
		\bibinfo {author} {\bibfnamefont {H.-C.}\ \bibnamefont {Po}},\ and\ \bibinfo
		{author} {\bibfnamefont {R.-B.}\ \bibnamefont {Liu}},\ }\bibfield  {title}
	{\bibinfo {title} {Phase transitions in the complex plane of physical
			parameters},\ }\href {https://doi.org/10.1038/srep05202} {\bibfield
		{journal} {\bibinfo  {journal} {Sci. Rep.}\ }\textbf {\bibinfo {volume}
			{4}},\ \bibinfo {pages} {5202} (\bibinfo {year} {2014})}\BibitemShut
	{NoStop}%
	\bibitem [{\citenamefont {Deger}\ \emph {et~al.}(2020)\citenamefont {Deger},
		\citenamefont {Brange},\ and\ \citenamefont {Flindt}}]{Deger2020}%
	\BibitemOpen
	\bibfield  {author} {\bibinfo {author} {\bibfnamefont {A.}~\bibnamefont
			{Deger}}, \bibinfo {author} {\bibfnamefont {F.}~\bibnamefont {Brange}},\ and\
		\bibinfo {author} {\bibfnamefont {C.}~\bibnamefont {Flindt}},\ }\bibfield
	{title} {\bibinfo {title} {{Lee-Yang theory, high cumulants, and
				large-deviation statistics of the magnetization in the Ising model}},\ }\href
	{https://doi.org/10.1103/PhysRevB.102.174418} {\bibfield  {journal} {\bibinfo
			{journal} {Phys. Rev. B}\ }\textbf {\bibinfo {volume} {102}},\ \bibinfo
		{pages} {174418} (\bibinfo {year} {2020})}\BibitemShut {NoStop}%
	\bibitem [{\citenamefont {Touchette}(2009)}]{Touchette2009}%
	\BibitemOpen
	\bibfield  {author} {\bibinfo {author} {\bibfnamefont {H.}~\bibnamefont
			{Touchette}},\ }\bibfield  {title} {\bibinfo {title} {The large deviation
			approach to statistical mechanics},\ }\href
	{https://doi.org/10.1016/j.physrep.2009.05.002} {\bibfield  {journal}
		{\bibinfo  {journal} {Phys. Rep.}\ }\textbf {\bibinfo {volume} {478}},\
		\bibinfo {pages} {1} (\bibinfo {year} {2009})}\BibitemShut {NoStop}%
	\bibitem [{\citenamefont {Ashhab}(2013)}]{sahel13}%
	\BibitemOpen
	\bibfield  {author} {\bibinfo {author} {\bibfnamefont {S.}~\bibnamefont
			{Ashhab}},\ }\bibfield  {title} {\bibinfo {title} {Superradiance transition
			in a system with a single qubit and a single oscillator},\ }\href
	{https://doi.org/10.1103/PhysRevA.87.013826} {\bibfield  {journal} {\bibinfo
			{journal} {Phys. Rev. A}\ }\textbf {\bibinfo {volume} {87}},\ \bibinfo
		{pages} {013826} (\bibinfo {year} {2013})}\BibitemShut {NoStop}%
	\bibitem [{\citenamefont {Hwang}\ \emph {et~al.}(2015)\citenamefont {Hwang},
		\citenamefont {Puebla},\ and\ \citenamefont {Plenio}}]{Hwang:2015}%
	\BibitemOpen
	\bibfield  {author} {\bibinfo {author} {\bibfnamefont {M.-J.}\ \bibnamefont
			{Hwang}}, \bibinfo {author} {\bibfnamefont {R.}~\bibnamefont {Puebla}},\ and\
		\bibinfo {author} {\bibfnamefont {M.~B.}\ \bibnamefont {Plenio}},\ }\bibfield
	{title} {\bibinfo {title} {{Quantum Phase Transition and Universal Dynamics
				in the Rabi Model}},\ }\href {https://doi.org/10.1103/PhysRevLett.115.180404}
	{\bibfield  {journal} {\bibinfo  {journal} {Phys. Rev. Lett.}\ }\textbf
		{\bibinfo {volume} {115}},\ \bibinfo {pages} {180404} (\bibinfo {year}
		{2015})}\BibitemShut {NoStop}%
	\bibitem [{\citenamefont {Hwang}\ \emph {et~al.}(2018)\citenamefont {Hwang},
		\citenamefont {Rabl},\ and\ \citenamefont {Plenio}}]{Hwang:2018}%
	\BibitemOpen
	\bibfield  {author} {\bibinfo {author} {\bibfnamefont {M.-J.}\ \bibnamefont
			{Hwang}}, \bibinfo {author} {\bibfnamefont {P.}~\bibnamefont {Rabl}},\ and\
		\bibinfo {author} {\bibfnamefont {M.~B.}\ \bibnamefont {Plenio}},\ }\bibfield
	{title} {\bibinfo {title} {{Dissipative phase transition in the open quantum
				Rabi model}},\ }\href {https://doi.org/10.1103/PhysRevA.97.013825} {\bibfield
		{journal} {\bibinfo  {journal} {Phys. Rev. A}\ }\textbf {\bibinfo {volume}
			{97}},\ \bibinfo {pages} {013825} (\bibinfo {year} {2018})}\BibitemShut
	{NoStop}%
	\bibitem [{\citenamefont {Zheng}\ \emph {et~al.}(2023)\citenamefont {Zheng},
		\citenamefont {Ning}, \citenamefont {Chen}, \citenamefont {L\"u},
		\citenamefont {Shen}, \citenamefont {Xu}, \citenamefont {Zhang},
		\citenamefont {Xu}, \citenamefont {Li}, \citenamefont {Xia}, \citenamefont
		{Wu}, \citenamefont {Yang}, \citenamefont {Miranowicz}, \citenamefont
		{Lambert}, \citenamefont {Zheng}, \citenamefont {Fan}, \citenamefont {Nori},\
		and\ \citenamefont {Zheng}}]{zheng24}%
	\BibitemOpen
	\bibfield  {author} {\bibinfo {author} {\bibfnamefont {R.-H.}\ \bibnamefont
			{Zheng}}, \bibinfo {author} {\bibfnamefont {W.}~\bibnamefont {Ning}},
		\bibinfo {author} {\bibfnamefont {Y.-H.}\ \bibnamefont {Chen}}, \bibinfo
		{author} {\bibfnamefont {J.-H.}\ \bibnamefont {L\"u}}, \bibinfo {author}
		{\bibfnamefont {L.-T.}\ \bibnamefont {Shen}}, \bibinfo {author}
		{\bibfnamefont {K.}~\bibnamefont {Xu}}, \bibinfo {author} {\bibfnamefont
			{Y.-R.}\ \bibnamefont {Zhang}}, \bibinfo {author} {\bibfnamefont
			{D.}~\bibnamefont {Xu}}, \bibinfo {author} {\bibfnamefont {H.}~\bibnamefont
			{Li}}, \bibinfo {author} {\bibfnamefont {Y.}~\bibnamefont {Xia}}, \bibinfo
		{author} {\bibfnamefont {F.}~\bibnamefont {Wu}}, \bibinfo {author}
		{\bibfnamefont {Z.-B.}\ \bibnamefont {Yang}}, \bibinfo {author}
		{\bibfnamefont {A.}~\bibnamefont {Miranowicz}}, \bibinfo {author}
		{\bibfnamefont {N.}~\bibnamefont {Lambert}}, \bibinfo {author} {\bibfnamefont
			{D.}~\bibnamefont {Zheng}}, \bibinfo {author} {\bibfnamefont
			{H.}~\bibnamefont {Fan}}, \bibinfo {author} {\bibfnamefont {F.}~\bibnamefont
			{Nori}},\ and\ \bibinfo {author} {\bibfnamefont {S.-B.}\ \bibnamefont
			{Zheng}},\ }\bibfield  {title} {\bibinfo {title} {{Observation of a
				Superradiant Phase Transition with Emergent Cat States}},\ }\href
	{https://doi.org/10.1103/PhysRevLett.131.113601} {\bibfield  {journal}
		{\bibinfo  {journal} {Phys. Rev. Lett.}\ }\textbf {\bibinfo {volume} {131}},\
		\bibinfo {pages} {113601} (\bibinfo {year} {2023})}\BibitemShut {NoStop}%
	\bibitem [{\citenamefont {Ge}\ \emph {et~al.}(2024)\citenamefont {Ge},
		\citenamefont {Fan},\ and\ \citenamefont {Nori}}]{Ge:2024}%
	\BibitemOpen
	\bibfield  {author} {\bibinfo {author} {\bibfnamefont {Z.-Y.}\ \bibnamefont
			{Ge}}, \bibinfo {author} {\bibfnamefont {H.}~\bibnamefont {Fan}},\ and\
		\bibinfo {author} {\bibfnamefont {F.}~\bibnamefont {Nori}},\ }\bibfield
	{title} {\bibinfo {title} {Effective field theories and finite-temperature
			properties of zero-dimensional superradiant quantum phase transitions},\
	}\href {https://doi.org/10.1103/PhysRevResearch.6.023123} {\bibfield
		{journal} {\bibinfo  {journal} {Phys. Rev. Res.}\ }\textbf {\bibinfo {volume}
			{6}},\ \bibinfo {pages} {023123} (\bibinfo {year} {2024})}\BibitemShut
	{NoStop}%
	\bibitem [{\citenamefont {Lipkin}\ \emph {et~al.}(1965)\citenamefont {Lipkin},
		\citenamefont {Meshkov},\ and\ \citenamefont {Glick}}]{Lipkin:1965}%
	\BibitemOpen
	\bibfield  {author} {\bibinfo {author} {\bibfnamefont {H.~J.}\ \bibnamefont
			{Lipkin}}, \bibinfo {author} {\bibfnamefont {N.}~\bibnamefont {Meshkov}},\
		and\ \bibinfo {author} {\bibfnamefont {A.~J.}\ \bibnamefont {Glick}},\
	}\bibfield  {title} {\bibinfo {title} {Validity of many-body approximation
			methods for a solvable model},\ }\href
	{https://doi.org/10.1016/0029-5582(65)90862-x} {\bibfield  {journal}
		{\bibinfo  {journal} {Nucl. Phys.}\ }\textbf {\bibinfo {volume} {62}},\
		\bibinfo {pages} {188} (\bibinfo {year} {1965})}\BibitemShut {NoStop}%
	\bibitem [{\citenamefont {Meshkov}\ \emph {et~al.}(1965)\citenamefont
		{Meshkov}, \citenamefont {Glick},\ and\ \citenamefont
		{Lipkin}}]{Meshkov:1965}%
	\BibitemOpen
	\bibfield  {author} {\bibinfo {author} {\bibfnamefont {N.}~\bibnamefont
			{Meshkov}}, \bibinfo {author} {\bibfnamefont {A.~J.}\ \bibnamefont {Glick}},\
		and\ \bibinfo {author} {\bibfnamefont {H.}~\bibnamefont {Lipkin}},\
	}\bibfield  {title} {\bibinfo {title} {Validity of many-body approximation
			methods for a solvable model},\ }\href
	{https://doi.org/10.1016/0029-5582(65)90863-1} {\bibfield  {journal}
		{\bibinfo  {journal} {Nucl. Phys.}\ }\textbf {\bibinfo {volume} {62}},\
		\bibinfo {pages} {199} (\bibinfo {year} {1965})}\BibitemShut {NoStop}%
	\bibitem [{\citenamefont {Glick}\ \emph {et~al.}(1965)\citenamefont {Glick},
		\citenamefont {Lipkin},\ and\ \citenamefont {Meshkov}}]{Glick:1965}%
	\BibitemOpen
	\bibfield  {author} {\bibinfo {author} {\bibfnamefont {A.~J.}\ \bibnamefont
			{Glick}}, \bibinfo {author} {\bibfnamefont {H.~J.}\ \bibnamefont {Lipkin}},\
		and\ \bibinfo {author} {\bibfnamefont {N.}~\bibnamefont {Meshkov}},\
	}\bibfield  {title} {\bibinfo {title} {Validity of many-body approximation
			methods for a solvable model},\ }\href
	{https://doi.org/10.1016/0029-5582(65)90864-3} {\bibfield  {journal}
		{\bibinfo  {journal} {Nucl. Phys.}\ }\textbf {\bibinfo {volume} {62}},\
		\bibinfo {pages} {211} (\bibinfo {year} {1965})}\BibitemShut {NoStop}%
	\bibitem [{\citenamefont {Kopylov}\ \emph {et~al.}(2017)\citenamefont
		{Kopylov}, \citenamefont {Schaller},\ and\ \citenamefont
		{Brandes}}]{Kopylov:2017}%
	\BibitemOpen
	\bibfield  {author} {\bibinfo {author} {\bibfnamefont {W.}~\bibnamefont
			{Kopylov}}, \bibinfo {author} {\bibfnamefont {G.}~\bibnamefont {Schaller}},\
		and\ \bibinfo {author} {\bibfnamefont {T.}~\bibnamefont {Brandes}},\
	}\bibfield  {title} {\bibinfo {title} {{Nonadiabatic dynamics of the excited
				states for the Lipkin-Meshkov-Glick model}},\ }\href
	{https://doi.org/10.1103/PhysRevE.96.012153} {\bibfield  {journal} {\bibinfo
			{journal} {Phys. Rev. E}\ }\textbf {\bibinfo {volume} {96}},\ \bibinfo
		{pages} {012153} (\bibinfo {year} {2017})}\BibitemShut {NoStop}%
\end{thebibliography}
\end{document}